\documentclass[a4paper,fleqn,usenatbib]{mnras}

\usepackage{natbib}

\usepackage[T1]{fontenc}
\usepackage{ae,aecompl}

\usepackage{graphicx}	
\usepackage{amsmath}	
\usepackage{amssymb}	

\renewcommand{\vec}[1]{ {\bf\it #1} }

\title[Simulating thermal and kinetic black hole feedback]
{Simulating galaxy formation with black hole driven thermal and kinetic feedback}

\author[R. Weinberger et al.]{%
Rainer Weinberger$^{1}$\thanks{E-mail: rainer.weinberger@h-its.org},
Volker Springel$^{1,2}$,
Lars Hernquist$^{3}$, 
Annalisa Pillepich$^{3,4}$,
\newauthor
Federico Marinacci$^{5}$, 
R\"udiger Pakmor$^{1}$, 
Dylan Nelson$^{6}$, 
Shy Genel$^{7}$\thanks{Hubble Fellow},
\newauthor
Mark Vogelsberger$^{5}$,
Jill Naiman$^{3}$,
Paul~Torrey$^{5,8}$\thanks{Hubble Fellow}
\vspace*{0.2cm}\\
$^{1}$Heidelberg Institute for Theoretical Studies, Schloss-Wolfsbrunnenweg 35, 69118 Heidelberg, Germany\\
$^{2}$Zentrum f\"ur Astronomie der Universit\"at Heidelberg, ARI, M\"onchhofstr. 12-14, 69120 Heidelberg, Germany\\
$^{3}$Harvard-Smithsonian Center for Astrophysics, 60 Garden Street, Cambridge, MA 02138, USA\\
$^{4}$Max-Planck-Institut f\"ur Astronomie, K\"onigstuhl 17, 69117 Heidelberg, Germany\\
$^{5}$Department of Physics, Kavli Institute for Astrophysics and Space Research, MIT, Cambridge, MA 02139, USA\\
$^{6}$Max-Planck-Institut
f\"{u}r Astrophysik, Karl-Schwarzschild-Stra\ss{}e 1, 85740 Garching
bei M\"{u}nchen, Germany\\
$^{7}$Department of Astronomy, Columbia University, 550 West 120th Street, New York, NY 10027, USA\\
$^{8} $TAPIR, Mailcode 350-17, California Institute of Technology, Pasadena, CA 91125, USA \\
}

\date{Accepted XXX. Received YYY; in original form ZZZ}

\pubyear{2016}

\begin{document}
\label{firstpage}
\pagerange{\pageref{firstpage}--\pageref{lastpage}}
\maketitle
\begin{abstract}
  The inefficiency of star formation in massive elliptical galaxies is widely believed to be caused by the interactions of an active galactic nucleus (AGN) with the surrounding gas. Achieving a sufficiently rapid reddening of moderately massive galaxies without expelling too many baryons has however proven difficult for hydrodynamical simulations of galaxy formation, prompting us to explore a new model for the accretion and feedback effects of supermassive black holes.  For high accretion rates relative to the Eddington limit, we assume that a fraction of the accreted rest mass energy heats the surrounding gas thermally, similar to the `quasar mode' in previous work.  For low accretion rates, we invoke a new, pure kinetic feedback model that imparts momentum to the surrounding gas in a stochastic manner.  These two modes of feedback are motivated both by theoretical conjectures for the existence of different types of accretion flows as well as recent observational evidence for the importance of kinetic AGN winds in quenching galaxies.  We find that a large fraction of the injected kinetic energy in this mode thermalizes via shocks in the surrounding gas, thereby providing a distributed heating channel. In cosmological simulations, the resulting model produces red, non star-forming massive elliptical galaxies, and achieves realistic gas fractions, black hole growth histories and thermodynamic profiles in large haloes. 

\end{abstract}




\section{Introduction}

In simulations of galaxy formation, feedback from active galactic nuclei (AGNs) is the most commonly invoked physical mechanism to explain the suppression of star formation in massive galaxies, and the observed correlations between black hole masses and properties of their host galaxies. In particular, feedback from luminous quasars has been suggested to limit black hole growth and star formation during mergers at high redshift \citep{2005Natur.433..604D, 2005MNRAS.361..776S, 2006ApJS..163....1H,2010MNRAS.406L..55D, 2014MNRAS.442..440C}. Interacting galaxies trigger a redistribution of angular momentum and thus gas inflows into the nuclear region of galaxies \citep{1989Natur.340..687H, 1996ApJ...471..115B, 1996ApJ...464..641M}. These gas inflows then generate a cascade of gravitational instabilities \citep{2010MNRAS.407.1529H, 2015MNRAS.446.2468E}, through which the supermassive black hole (SMBH) is fuelled and a fraction of the gravitational binding energy is released. This energy is sufficient to lower the star formation rate by several orders of magnitude \citep{2005Natur.433..604D}. However, it is not yet clear whether the released energy has a lasting effect on the whole galaxy and its star formation rate, or just affects the innermost regions \citep{2011MNRAS.412.1341D,2015ApJ...800...19R}. 

{By applying semi-analytic modelling, \citet{2006MNRAS.365...11C} pointed out that  `radio-mode' feedback, which provides an efficient source of energy in systems with hot, hydrostatic atmospheres, can simultaneously explain the low mass drop-out rate in cooling flows, the exponential cutoff at the bright end of the galaxy luminosity function and the increased mean stellar age in massive elliptical galaxies.
\citet{2006MNRAS.370..645B} used a similar approach in their semi-analytic model.}
 \citet{2007MNRAS.380..877S} presented a unified sub-resolution model with energy input from both quasars and radio-mode feedback in hydrodynamical simulations and applied it to galaxy cluster formation. In this model, the second mode of feedback is active once the black hole accretion rate relative to the Eddington limit, $\dot{M}_\text{BH} / \dot{M}_\text{Edd}$, drops below a given value. The feedback energy injection is modelled by heating up spherical bubbles of gas in galaxy haloes, mimicking the observed radio lobes in galaxy clusters.

There are various implementations of `quasar mode' feedback in the literature.  \citet{2011MNRAS.412.1341D,2012MNRAS.420.2221D} use feedback from radiation pressure from luminous AGN, modelled by depositing momentum in surrounding simulation particles in idealized mergers. \citet{2012ApJ...754..125C,2014MNRAS.442..440C,2015MNRAS.449.4105C} included mechanical and thermal energy and pressure from X-rays in their AGN feedback prescription and studied the effect on idealized mergers of disc galaxies and in cosmological ``zoom'' simulations of elliptical galaxies while \citet{2013MNRAS.431.2513W} performed a comparative study of these AGN models in merger simulations.

Likewise, many different approaches for `radio mode' activity have been taken, often using bipolar outflows in idealized simulations of hydrostatic haloes \citep{2002MNRAS.332..271R,2003MNRAS.339..353B,2004MNRAS.348.1105O,2004ApJ...611..158R,2005A&A...429..399Z,2006ApJ...643..120B,2007MNRAS.380L..67B,2007MNRAS.376.1547C,2007ApJ...656L...5S,2009MNRAS.395..228S,2011MNRAS.411..349G, 2011MNRAS.415.1549G,2012MNRAS.424..190G,2014ApJ...789...54L,2014ApJ...789..153L,2015ApJ...811...73L,2016MNRAS.455.2139H,2016ApJ...818..181Y,2016arXiv160501725Y}, or in cosmological simulations \citep{2010MNRAS.409..985D,2012MNRAS.420.2662D,2016arXiv160603086D}. These methods assume that quenching is caused by the energy that is released from collimated jets and their associated radio lobes, which can be found in massive systems \citep{2006MNRAS.373..959D}. However, \citet{2016arXiv160303674M} show that these kinetic feedback implementations have a different impact in an idealized galaxy cluster setup compared to pure thermal injection.

The extensive body of literature on coupled AGN-galaxy evolution  \citep[including][among others]{2004ApJ...600..580G,2005MNRAS.358L..16K,2007MNRAS.380..877S,2008ApJ...676...33D,2008ApJS..175..356H,2008ApJS..175..390H,2008MNRAS.385..161O,2008MNRAS.391..481S,2009MNRAS.398...53B,2010ApJ...717..708C,2010MNRAS.406L..55D,2011MNRAS.414..195T,2012MNRAS.420.2662D,2013arXiv1312.0598R,2014MNRAS.442.2304H,2014MNRAS.442..440C,2015MNRAS.450.1349K,2015ARA&A..53...51S,2015MNRAS.448.1504S,2016MNRAS.460.3925T}
has recently been complemented by a new generation of
high-resolution cosmological simulations of galaxy formation in large volumes, such as Eagle \citep{2015MNRAS.446..521S} and Illustris \citep{2014Natur.509..177V}. 
The corresponding implementations for black hole feedback in massive galaxies (in Illustris the radio-mode, while Eagle does not distinguish between modes) gather energy up to a predetermined threshold value, which parametrizes its burstiness and inject it instantaneously as thermal energy \citep[see][for Illustris and Eagle, respectively]{2007MNRAS.380..877S,2009MNRAS.398...53B}.

While the Illustris simulation -- which forms the starting point of our work -- has been remarkably successful in matching a wide range of galaxy properties, its results are in tension with a number of properties of observed haloes and galaxies.  An important discrepancy arising from the AGN feedback model is the gas fraction of groups of galaxies and poor clusters, which is substantially too low in Illustris \citep{2014MNRAS.445..175G}. At the same time, the stellar masses of the central galaxies in the simulated systems are too high. Employing a yet higher feedback efficiency of the BH radio mode to suppress star formation further would expel even more gas, and hence does not represent a viable solution. Alternatively, as part of our study, we made numerous attempts to improve the impact of the bubble model by adopting different choices for the parameters or by adding non-thermal pressure support in the form of magnetic fields, but without success.  We therefore conclude that the particular AGN feedback model in Illustris is disfavoured, and a more radical change is in order.

This suggestion is supported by recent observational findings about the possible importance of kinetic winds driven during BH accretion. For example, \citet{Cheung2016} find bisymmetric emission features in the centres of quiescent galaxies of stellar mass around $2 \times 10^{10}\,{\rm M}_\odot$, from which they infer the presence of centrally driven winds in typical quiescent galaxies that host low-luminosity active nuclei. They show that such `red geyser' galaxies are very common at this mass scale, and that the energy input from the low activity of the SMBHs of these galaxies is capable of driving the observed winds, which contain sufficient mechanical energy to suppress star formation. This appears to be a feedback channel that is distinct from the radio galaxies at the centres of clusters, but as it affects many more galaxies at lower mass scales, it could well be more important for global galaxy evolution. {Recently, \citet{2016arXiv160702507P} found hot, AGN-driven outflows in post-merger galaxies, using the single-mode thermal AGN feedback model of \citet{2016arXiv160702151T}}. Interestingly {however}, \citet{2014ApJ...796....7G} and \citet{2014ApJ...787...38F} have discovered wide-spread, powerful AGN-driven outflows in the majority ($\sim 70\%$) of massive $z \sim 1-2$ star-forming galaxies. Because this phenomenon is so common, {it likely arises from low-luminosity AGN with low Eddington ratios  and thus appears consistent with a kinetic wind mode.}  Also, theoretically there is good motivation for hot coronal winds from BH accretion flows. For example, \citet{2014ARA&A..52..529Y} discuss such a scenario, which can be viewed as a small-scale version of the jet model of \citet{Blandford1977}.

The motivation of our  work is therefore to develop a revised model for black hole growth and feedback that takes these considerations into account. It is important to realize that the relevant time and length scales of the detailed black hole physics are by far not resolved in cosmological simulations.  Hence, the corresponding feedback models can only be implemented as so-called sub resolution treatments that mimic the net effect of feedback on resolved scales. Besides the theoretical uncertainties involved, this approach comes with the drawback that the behaviour of the models can vary between different numerical methods, because the scales at which the gas state is affected by the subgrid treatment are only marginally resolved. This is demonstrated for example in \citet{2015MNRAS.452..575S} for the bubble heating model of \citet{2007MNRAS.380..877S}. We thus also aim to take recent improvements in the accuracy of the hydrodynamical modelling into account \citep{2012MNRAS.423.2558B, 2012MNRAS.425.2027K, 2012MNRAS.424.2999S, 2012MNRAS.425.3024V, 2016MNRAS.455.1134P}.

The model presented here conjectures two modes of feedback from AGN in thermal and kinetic form, and in this sense is similar to \citet{2012MNRAS.420.2662D}. While the kinetic part of their model is inspired by the sub-relativistic jet simulations of \citet{2004MNRAS.348.1105O}, our approach does not directly aim to represent jets from AGNs that act on marginally resolved scales. Rather we assume that the physical mechanisms that provide energy and momentum transport from black holes to their surroundings are reasonably efficient, and that their impact on large scales can be captured by depositing energy and momentum in small regions around halo centres. This approach does not address the microphysics of the origin of AGN feedback but aims to arrive at a robust parametrization of the effects of black holes on galaxy and galaxy cluster formation even at coarse resolution.

In what follows, we present a new model for SMBH growth and AGN feedback in cosmological simulations of structure formation implemented in the moving-mesh magnetohydrodynamics code {\small AREPO} \citep{2010MNRAS.401..791S, 2011MNRAS.418.1392P,2016MNRAS.455.1134P}. In Section~\ref{sec:Model}, we describe the model and its free parameters. Because the main modification to previous works lies in feedback injection at low accretion rates, in Section~\ref{sec:Tests} we discuss idealized tests of how the energy couples in this mode to the gas. We then continue in Section~\ref{sec:CosmoSims} with an investigation of its impact on cosmological simulations of galaxy formation. Section~\ref{sec:parameters} is dedicated to a systematic exploration of the influence of the different model parameters on the results. Finally, we describe our findings and present our conclusions in Section~\ref{sec:Conclusion}. Appendix~\ref{app:wind} specifies, for definiteness, details of our supernova feedback model, and Appendix~\ref{app:convergence} discusses numerical resolution dependencies.


\section{Black Hole Model}
\label{sec:Model}

Modelling AGNs in cosmological simulations poses
several fundamental challenges.  First, the detailed physical mechanisms
of both accretion on to SMBHs
\citep{2010MNRAS.407.1529H,2011MNRAS.415.1027H,2013ApJ...770....5A,2013MNRAS.432.3401G, 2015ApJ...800..127A,2016arXiv160308007A, 2015MNRAS.454.3445C,2016arXiv160602729C, 2015MNRAS.446.2468E,2013arXiv1312.0598R}
and the AGN-gas interaction
\citep{2011MNRAS.418.1621H,2012MNRAS.425..438G,2014MNRAS.439.2903C,2014MNRAS.444.2355C,
  2015ApJ...800...19R, 2016arXiv160606281B,2015arXiv150405209H} are poorly understood, which makes it at present impossible to formulate a `correct' treatment for simulations, independent of their resolution. Secondly, the extreme dynamic range posed by the problem, where a comparatively tiny accretion region around the black hole influences an entire galaxy or even a galaxy cluster and the surrounding intergalactic medium, vastly exceeds the capabilities of current numerical techniques so that much of the physics on the smallest scales needs to be coarsely approximated with sub-resolution models.  Thirdly, the nonlinear nature of galaxy formation intimately couples black hole accretion with other aspects of feedback, chiefly the regulation of ordinary star formation \citep{2013MNRAS.428.2966P}. This makes it difficult to disentangle the impact of different astrophysical processes. While we first examine the behaviour of our model in well-defined idealized tests, we will primarily assess its performance through studies of its consequences in the full cosmological context.


Similar to \citet{2007MNRAS.380..877S}, we distinguish between states
of high and low accretion rates. This follows the theoretical
notion that there exist (at least) two physically distinct types
of accretion flows on to massive black holes \citep[e.g.][and references therein]{2014arXiv1410.8132B}: one at comparatively high
rates in a classic disc mode \citep{1973A&A....24..337S}, the other at lower rates in a more spherical and hotter accretion flow \citep{1976ApJ...204..187S,1977ApJ...214..840I}. These regimes have loosely been
identified with ``quasar'' and ``radio'' modes in previous simulation
work. The observed phenomenology of radio jets in galaxy clusters has
often been interpreted as providing the dominant source of feedback,
at least in the low-accretion radio mode regime \citep{2007ARA&A..45..117M}. This has also
motivated, e.g., the bubble heating model in
\citet{2007MNRAS.380..877S} that was applied in the Illustris
simulation and in other works. However, there are also theoretical
indications pointing to the existence of kinetic winds in the
low-accretion state \citep[][]{1999MNRAS.303..309I, 1999MNRAS.310.1002S, 2014ARA&A..52..529Y, 2015ApJ...804..101Y, 2015arXiv151003124B, 2015arXiv151008845S}. These would be difficult to observe but could constitute an even more important feedback mechanism than the radio jets themselves. A central motivation of our work is to test this idea by
replacing radio bubble feedback with a kinetic wind.

\subsection{Accretion mode}
\label{subsec:Accretion}

We follow previous work and use the Eddington ratio as the criterion for
deciding the accretion state of the black hole. Specifically, we
assume SMBHs to be in the high accretion state as long as their
Bondi-Hoyle-Lyttleton accretion rate $\dot{M}_\text{Bondi}$
\citep{1939PCPS...35..405H, 1944MNRAS.104..273B, 1952MNRAS.112..195B}
exceeds a fraction $\chi$ of the Eddington accretion
rate $\dot{M}_\text{Edd}$:
\begin{align}
\frac{ \dot{M}_{\text{Bondi}}}{\dot{M}_{\text{Edd}}} &\geq \chi, \label{eq:eddingtonfactor}
\end{align}
where
\begin{align}
\dot{M}_\text{Bondi} &= \frac{4 \pi G^2 M_\text{BH}^2 \rho}{c_s^3} ,\label{eq:Bondirate} \\
\dot{M}_\text{Edd} &= \frac{4 \pi G M_\text{BH} m_p}{\epsilon_r \sigma_{\rm T} c}. \label{eq:Eddingtonrate}
\end{align}
Here, $G$ denotes the gravitational constant, $c$ the vacuum speed of
light, $m_p$ the proton mass and $\sigma_{\rm T}$ the Thompson
cross-section. The factor $\epsilon_r$ is the radiative
accretion efficiency. 
$M_\text{BH}$ is the black hole mass, and $\rho$ and $c_s$ are the
density and sound speed\footnote{We use an effective sound speed, taking into account both thermal and magnetic signal propagation $c_\text{s}^2 = c_\text{s,therm}^2 + c_\text{A}^2$, where $c_\text{A} = \left(\vec{B}^2 / 4 \pi \rho\right)^{1/2}$ is the Alfven speed.} of the gas near the
black hole, respectively. They are obtained by averaging over a sphere with radius $h$ 
in a kernel-weighted fashion around the black hole such that the enclosed number of cells in this sphere is approximately equal to a prescribed number:
\begin{align}
n_\text{ngb} \approx \sum\limits_{i} \,\frac{4\, \pi\,h^3\,m_i}{3\,m_\text{baryon}} w(r_i).
\label{eq:nngb}
\end{align}
Here, $m_\text{baryon}$ is the target mass of a gas cell, i.e.~the gas mass resolution enforced by the refinement and derefinement operations of the hydrodynamic code, $n_\text{ngb}$ is the prescribed number of neighbouring cells in this averaging and $w(r)$ is an SPH weighting kernel.

A sensible value for $\chi$ is expected to lie in the range $\sim 0.001-0.1$, by analogy with X-ray binaries \citep[e.g.][]{2010MNRAS.403...61D}.  Previous works \citep{2007MNRAS.380..877S,2015MNRAS.452..575S} have employed a fixed value of $\chi$. Black holes at low redshift located in massive systems show clear signatures of being in a `radio' feedback state, \citep{2006MNRAS.373..959D} which indicates low Eddington ratios. However, as we will show in Section~\ref{sec:CosmoSims}, this does not occur in our simulations {\em unless} the black holes transition to the kinetic mode in the first place, which is not guaranteed. To favour this transition for the most massive black holes at late times (which tend to be found in the most massive haloes), we scale the threshold with black hole mass, \begin{align}
  \chi = \min\left[ \chi_{0}\left(\frac{M_\text{BH}}{10^8 \,\text{M}_\odot} \right)^\beta, 0.1\right]
\end{align}
with $\chi_{0}$ and $\beta$ as parameters. The pivot mass $10^8\,\text{M}_\odot$ is degenerate with $\chi_{0}$ and is therefore not set independently. We limit the threshold $\chi$ to a maximum of $0.1$ to always allow any black hole (including the most massive ones) to reach the high accretion state provided there is a large enough gas supply to fuel them. This would be expected for high redshift quasars that have very massive black holes.\footnote{Note that the volume of the simulations presented in this work (Section~\ref{sec:CosmoSims}) is too small to host these kind of objects.} For $\beta > 0$, our scaling makes it more difficult for low mass black holes to be in the kinetic mode, and vice versa. We expect this to support the occurrence of a rapid quenching transition in massive galaxies, and make it unlikely that low-mass galaxies will be strongly affected by kinetic feedback.  {The physics of the accretion mode transitions of SMBHs is poorly understood, making it difficult to parametrize it adequately in a coarse cosmological model. We have here opted for a heuristic model that is based on the only intrinsic black hole property we keep track of, the black hole mass, and which is selected pragmatically based on how well it reproduces observational trends.  We note that it appears physically plausible that there are systematic trends with black mass scale in the accretion mode transition, given that radiative cooling physics breaks the scale invariance.}

\subsection{Accretion estimate and seeding of black holes}
\label{subsec:model_seeding}

Note that in the above calculation of the Bondi accretion rate we omit a boost-factor $\alpha$ that was used in older models to account for the unresolved ISM structure. When the latter is treated with a sub-resolution model that prescribes a high mean thermal support and an effective pressure, the Bondi rate is artificially biased low, slowing down especially the early growth of black holes. The boost factor was primarily introduced in the older models to compensate for this problem {by ensuring that the Bondi growth time-scale for small mass seed black holes does not exceed the Hubble time}.  Since the actual accretion rate was however anyway limited to the Eddington rate, the latter is ultimately the governing rate for most of the growth. {Furthermore, applying a boost factor for massive black holes in the low accretion state, when their feedback generates a low-density, hot gas phase around them (that can be resolved, unlike the ISM), appears questionable.}  We therefore simplify our treatment by assuming that the black holes are always accreting at the pure Bondi rate, limited by the Eddington rate: \begin{align}
  \dot{M}_\text{BH} = \min\left( \dot{M}_\text{Bondi}, \dot{M}_\text{Edd} \right).
\end{align}

{We note that for massive black holes at late times, the accretion rate is self-regulated, thus an additional factor in the accretion rate estimate has no overall effect in this regime apart from systematically shifting the black hole masses, i.e.~here the boost factor is largely degenerate with the black hole masses reached. Only the early growth phase is strongly affected by the boost factor, but this phase depends sensitively on the black hole seed mass as well (see discussion below), and we use this dependence to make up for the omission of a boost factor. }

Note that \citet{2013MNRAS.436.3031V} lowered the accretion rate estimate by a factor of $(P_\text{ext} / P_\text{ref})^2$ whenever $P_\text{ext}<P_\text{ref}$. Here, $P_\text{ext}$ is the kernel-weighted gas pressure surrounding the black hole and $P_\text{ref}$ is a reference pressure \citep[][their equation 23]{2013MNRAS.436.3031V}. While this was used in the Illustris simulation, we omit such a factor in this work. We ran simulations both with and without this factor and found no significant difference in the properties presented in this work. However, as this serves as a protection against rare cases of overly heated, underdense regions in galaxy centres we plan to use it in future simulations that contain a larger sample of galaxies.

In our cosmological simulations, a black hole with mass $M_\text{seed}$ is placed at the centre of a halo whenever the on-the-fly friend-of-friends halo finder identifies a halo more massive than a threshold mass $M_\text{FOF}$ that does not yet contain a black hole.  We note that to offset a potentially sluggish growth of black holes at high redshift, one can resort to a slightly larger seed mass, which then produces a similar result as using a boost factor $\alpha$. In order to remain close to our previous models, we use this here and adopt a black hole seed mass of $8\times 10^5 \,h^{-1}\,\text{M}_\odot$ in our default model, which leads to a similarly fast growth at early times as our older models with $\alpha=100$. Given the significant theoretical uncertainties in the early growth of SMBHs \citep[e.g.][]{2010A&ARv..18..279V}, we consider the seed mass as a poorly constrained free parameter. {We note that there are other models for back hole seed formation in cosmological simulations \citep[e.g.][]{2011ApJ...742...13B, 2016arXiv160702151T} that use thresholds of local gas properties such as metallicity, density and temperature. However, we decided for a seeding prescription depending solely on halo mass because of its simplicity and numerical robustness.}

At the limited numerical resolution available in cosmological simulations, two-body discreteness effects and numerical $N$-body noise can displace black hole particles from halo centres. At the same time, the dynamical friction forces that should allow massive black holes to sink to the centres of dark matter haloes are not captured accurately by the simulation. To prevent black holes from artificially leaving the centres of haloes for long periods of time due to these effects, we resort to an ad hoc centring prescription designed to keep black holes very close to the potential minimum of their host dark matter haloes. To this end, at every global integration timestep (i.e.~when the longest timesteps occurring in the whole simulation are synchronized in the nested time integration scheme), we determine the minimum gravitational potential in a region around the BH containing the equivalent of 1000 mass resolution elements. The BH particle is then shifted to this potential minimum (if not at the location of the BH already, which frequently happens), and its velocity is set to the mean mass-weighted velocity of the region. The latter minimizes any motion of the BH with respect to the central region of the halo. This method robustly prevents haloes from losing their central black hole, and it further adopts a scenario in which BH binaries are assumed to merge promptly. {We use this approach here because of its numerical robustness and independence of resolution. However, there are more sophisticated treatments in the recent literature that use sub-resolution models for dynamical friction  \citep[e.g.][]{2013MNRAS.431.2513W,2014MNRAS.442.2304H,2016arXiv160702151T}. We aim to use such a scheme in future high-resolution extensions of the present model.}

\subsection{Feedback}
\label{subsec:feedback}

For the high accretion state, we calculate the liberated feedback energy as
\begin{align}
 \Delta \dot{E}_\text{high} = \epsilon_{\text{f,high}} \epsilon_{\text{r}} \dot{M}_\text{BH} \,c^2,
 \label{eq:feedbackenergy}
\end{align}
where $\dot{M}_\text{BH}$ is the estimated black hole mass accretion rate of the black hole with mass $M_\text{BH}$, $\epsilon_{\text{r}}$ is the radiative efficiency (i.e.~the
canonical 0.1-0.2 of the accreted rest-mass energy that is released in the accretion process and not vanishing in the black hole), while $\epsilon_{\text{f,high}}$ is the fraction of this energy that couples to the surrounding gas. 
For the low accretion state, the feedback energy is parameterized as
\begin{align}
\Delta \dot{E}_\text{low} = \epsilon_\text{f,kin} \dot{M}_\text{BH} c^2 .
\end{align}

Note that we use different coupling efficiencies, $\epsilon_\text{f,kin}$ and $\epsilon_\text{f,high}$, for the two modes, motivated by the different physical nature of the accretion modes; namely that the low accretion state is thought to be radiatively inefficient. We keep the coupling efficiency in the high accretion state at a constant value of $\epsilon_\text{f,high} = 0.1$, resulting in an overall efficiency $\epsilon_\text{r} \epsilon_\text{f,high} = 0.02$, while we set a maximum value of $\epsilon_\text{f,kin} = 0.2$ in the low accretion mode, which assumes that the released rest-mass energy appears primarily in kinetic outflows. {We note that our choice of $\epsilon_\text{f,kin} = 0.2$ is within the physically plausible range, depending on the underlying physical mechanism.  If, for example, the energy is delivered by small-scale jets produced by the Blandford-Znajek mechanism, the jet energy can be a factor of $10$ or more larger than our adopted value for $\epsilon_\text{f,kin}$ because of black hole spin \citep{2014ARA&A..52..529Y}. In the opposite case, for non-spinning black holes, small-scale simulations of accretion \citep{2015ApJ...804..101Y} provide a theoretical lower limit to $\epsilon_\text{f,kin}$ of about $10^{-3}$.}

To protect against a potential runaway of the kinetic feedback mode that may drive the density to ever lower values \citep[see also section 2.6.2 of][for further discussion]{2013MNRAS.436.3031V}, we conjecture that at very low densities the coupling efficiency $\epsilon_\text{f,kin}$ eventually becomes weak. For simplicity, we assume that such a weakening occurs below a density $f_\text{thresh}\rho_\text{SFthresh}$, where $f_\text{thresh}$ is a free parameter and $\rho_\text{SFthresh}$ is the density threshold for star formation. If the surrounding density $\rho$ drops below this value we reduce the coupling proportional to density. This then formally corresponds to a variable coupling efficiency in the low accretion state, \begin{align}
  \epsilon_\text{f,kin} = \min\left( \frac{\rho}{f_\text{thresh} \rho_\text{SFthresh}} , 0.2\right).
\end{align}
{Our standard value for this prescription is
$f_\text{thresh} = 0.05$, and we will show in Section~\ref{sec:parameters} that the exact value of $\epsilon_\text{f,kin}$ has hardly any impact on galaxy properties.}

In the high accretion state, we inject the feedback as pure thermal energy in a small local environment around the black hole, as in \citet{2005MNRAS.361..776S} and our subsequent work \citep[including][as well as Illustris]{2013MNRAS.436.3031V}, while using our new model of kinetic feedback in the low accretion state. In the latter case, we inject the energy as pure kinetic energy. Unlike in the high accretion state, we hence input momentum but no immediate thermal energy to the gas.  Technically, we inject both forms of feedback in a kernel-weighted manner into a prescribed number of gas neighbouring the BH, as determined by equation.~(\ref{eq:nngb}). This region is identical for imparting feedback and the calculation of the gas properties used in the accretion estimate.

Because we cannot spatially resolve small-scale jets and the accretions flows in our cosmological simulations, we add the momentum in a random direction. We have found that this approach is most robust for avoiding possible numerical artefacts that can be produced at poor resolution by more elaborate approaches for adding the momentum. For example, one may impart the momentum in a spherically symmetric fashion, radially away from the black hole, with zero total momentum (as vector sum) added per injection event. However, this can produce an artificial suppression of the gas density at the position of the black hole at the resolution we achieve here. Similarly, a biconical injection in opposite directions at the position of the black hole can create artificially depressed gas densities unless the `jets' are well enough resolved. We therefore prefer random injection directions that change for every injection event, which we found to be least resolution dependent. In this case detailed energy and momentum conservation is only obtained as a time average over the injection events.

Specifically, for an available kinetic feedback energy $\Delta E$, we kick each gas cell $j$ in the feedback region by 
\begin{align}
  \Delta \vec{p}_j = m_j \sqrt{\frac{2\, \Delta E\, w(\vec{r}_j)}{\rho}} \, \vec{n} ,
\end{align}
where $\Delta\vec{p}_j$ is the change in momentum of gas cell $j$, $m_j$ denotes its mass, and $\vec{r}_j$ is the distance vector from the black hole to the respective cell. The factor $\vec{n}$ is the unit vector in a randomly chosen injection direction, $w(\vec{r}_j)$ the value of the smoothing kernel, and $\rho$ is the density estimate of the surrounding gas, as described in Section~\ref{subsec:Accretion}. 
The total momentum injection per feedback event is thus
\begin{align}
  \vec{p}_\text{inj} &= \sum\limits_j m_j \sqrt{\frac{2\, \Delta E\, w(\vec{r}_j)}{\rho}} \, \vec{n},
\end{align}
and the corresponding change in total energy of the gas (relative to the lab frame) is
\begin{align}
 E_\text{inj} &= \Delta E + \sum\limits_j \left(\vec{p}_j \cdot \vec{n}\right) \sqrt{\frac{2\,\Delta E\, w(\vec{r}_j)}{\rho}},
\end{align}
where $\vec{p}_j$ is the momentum of cell $j$ before the injection event.

For a single injection event, this violates strict momentum conservation and will generally not increase the total energy by precisely $\Delta E$. However, the average over many injection events leads to the desired energy injection and assures momentum conservation (i.e. $\left<\vec{p}_\text{inj} \right> = 0$), as the injection direction $\vec{n}$ is randomly chosen for each injection event and does not correlate with the flow direction of the surrounding gas. To make the occurrence of these injection events independent of the timestepping, and also to make them powerful enough individually, we discretize the kinetic feedback mode by imposing a minimum energy that needs to accumulate in the kinetic accretion mode before the feedback is released. This is similar to the approach adopted in Illustris and Eagle for the BH feedback in large haloes.

In this work, we choose to parametrize the adopted energy threshold for the kinetic feedback in terms of a fiducial energy computed from the mass of the feedback region and the surrounding dark matter velocity dispersion.  This identifies an energy per unit mass that is tied to the virial temperature of the halo. In fact, we could also construct the energy scale from the temperature of the surrounding gas. But the latter can be affected strongly by local cooling or previous feedback events, hence we prefer to use the dark matter velocity dispersion for increased robustness. We note that the velocity dispersion is also used for our supernova-driven wind feedback from star formation, which we adopt from Illustris in only a slightly modified form (see Appendix~\ref{app:wind}).  We parametrize the kinetic feedback threshold by 
\begin{align}
  E_{\rm inj,min} = f_\text{re} \frac{1}{2}\,\sigma_{\rm DM}^2\, m_\text{enc}
  \label{eq:Einj_min}
\end{align}
where $\sigma_{\rm DM}$ is the 1D dark matter velocity dispersion, $m_\text{enc}$ is the gas mass in the feedback region, and $f_\text{re}$ is a free parameter that specifies the burstiness and thus the frequency of the reorientation of the kinetic feedback. If a larger value is chosen for $f_\text{re}$, fewer feedback events occur, but they are individually stronger. {Choosing this scaling is partly numerically motivated, as it ensures that the resulting shocks are strong enough to be accurately captured by our finite-volume scheme. Without this threshold, low-luminosity black holes would drive very weak flows that would thermalize mainly via numerical dissipation effects, which is clearly undesirable.
  Part of the motivation is also physical, because this scaling ensures that the specific energy of the wind does not significantly exceed the specific binding energy of the halo, and thus should not unbind a large amount of gas or overly disturb the thermodynamic state of the intrahalo gas.}


\section{Kinetic wind dissipation tests}
\label{sec:Tests}

\begin{figure}
 \centering
 \includegraphics{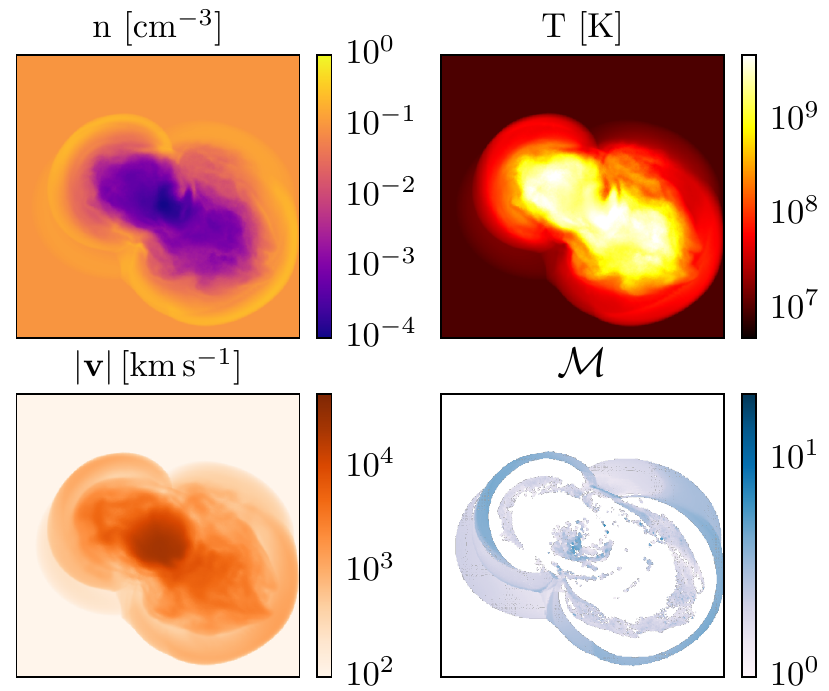}
 \caption{Thin projection ($5$ kpc in depth, $25$ kpc on a side) of
   the $256^3$, $n=10^{-1}\,\text{cm}^{-3}$, $T=10^7\,\text{K}$
   simulation after $5\,\text{Myr}$ of evolution. The panels show
   volume weighted density (top left), volume-weighted temperature
   (top right), absolute velocity (bottom left), and energy
   dissipation weighted Mach number (bottom right).}
\label{fig:idealised_projection}
\end{figure}

\begin{figure}
 \centering
 \includegraphics{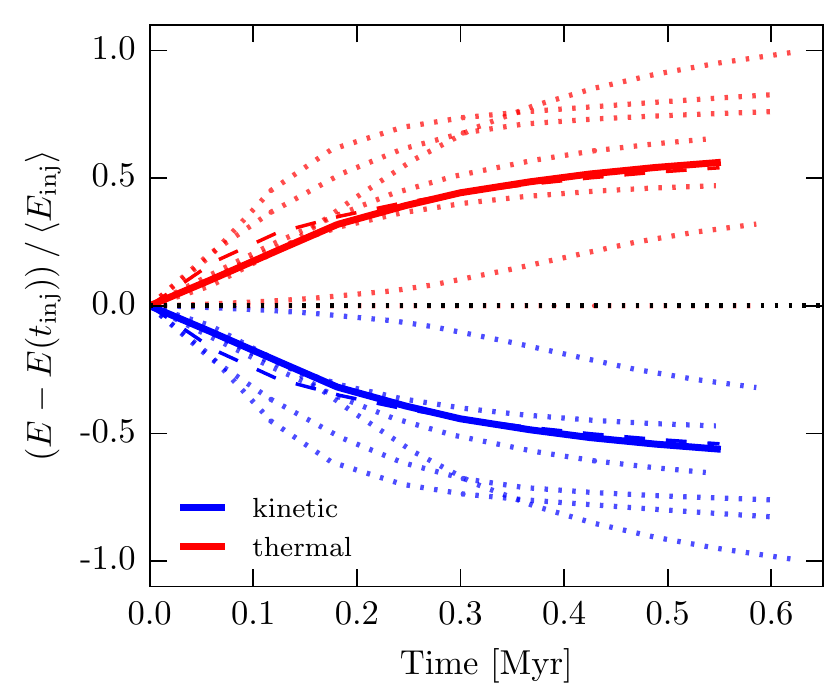}
 \caption{Evolution of the different energy components after kinetic energy injection. The dotted
   lines show individual injection events, the solid line their
   average, both in the simulation initially with $32^3$ cells. The
   dashed line shows the average of the high resolution test with
   $256^3$ initial cells. On average, half of the feedback energy that was initially in kinetic form is thermalized after $0.5$ Myr. This behaviour is converged at the resolution of cosmological simulations.}
\label{fig:E_time_periodic}
\end{figure}

\begin{figure}
 \centering
 \includegraphics{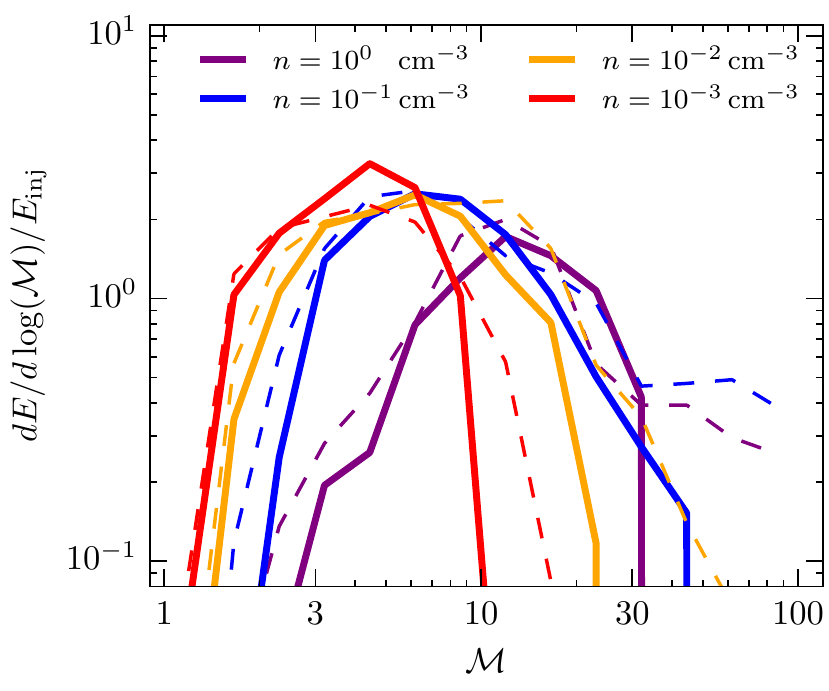}
 \caption{Energy dissipation as a function of shock Mach number
   $\mathcal{M}$ summed up over a simulation time of
   $5\,\text{Myr}$. The different colours denote different isobaric
   variations of the gas from relatively cool, dense
   ($1 \,\text{cm}^{-3}$, $T= 10^{6}\,\text{K}$) to hot, dilute
   ($\rho=10^{-3}\,\text{cm}^{-3}$, $T= 10^{9}\, \text{K}$). The solid
   lines show the simulation with an initial grid of $32^3$ cells,
   comparable to the resolution of cosmological simulations, while the
   dashed lines indicate simulations with $256^3$ cells to show the
   convergence of the analysis.}
\label{fig:M_Statistic_rho}
\end{figure}

To examine the dissipation mechanisms of the kinetic feedback model, we use idealized test simulations in a cubic box with constant density, temperature and pressure and a side length of $25\, \text{kpc}$. The fiducial values for density and temperature are $n = 10^{-1}\, \text{cm}^{-3}$ and $T = 10^{7}\, \text{K}$. We place a black hole at the centre and inject energy at a fixed rate of $10^{45}\,\text{erg}\,\text{s}^{-1}$ in the kinetic mode. We run the simulations for $5 \,\text{Myr}$, only solving the equations of hydrodynamics, switching off self-gravity, gas cooling and all galaxy formation sub-grid prescriptions such as star-formation and feedback, metal enrichment, black hole seeding, etc. We run the simulations at two different resolutions: $32^3$ initial cells, which roughly corresponds to the resolution of cosmological simulations (for $\rho = 10^{-1}\, \text{cm}^{-3}$, the average mass of a gas cell is $7\times 10^{5}\,\text{M}_\odot$), and $256^3$, to show the convergence properties.

Unlike in cosmological simulations, where we keep the number of neighbours in the feedback injection region roughly constant, we here fix the radius of the sphere in which the feedback is injected. We have tested both approaches and found that there is no substantial difference, except that tying the injection region to the number of neighbouring gas cells (equation.~\ref{eq:nngb}) leads in this particular setup -- in which self-regulation is disabled -- to a slowly growing injection region, as the gas around the black hole is heated up by previous feedback events.  To promote a clean study of the impact of the kinetic pulses on the gas, we prefer to keep the feedback injection region fixed to a sphere of $3.5\,\text{kpc}$ radius around the black hole. We ensure that there is always a sufficient number of cells in this region by setting a maximum volume per cell, above which they are refined\footnote{Note that in cosmological simulations, we do not impose such a volume limit.}. The physical size of the feedback injection region is kept the same for the $256^3$ simulations, allowing a study of discretization effects in the gas. We note, however, that in cosmological simulations there are additional resolution dependences such as the scaling of the feedback injection region, which is discussed in Appendix~\ref{app:convergence}. As there is no dark matter in the present test simulations, we replace equation~(\ref{eq:Einj_min}) with \begin{align}
  E_{\rm inj,min} = f_\text{re}\,u_\text{init}\, m_\text{enc},
\end{align}
where $u_\text{init}$ is the initial specific thermal energy. This means that we are assuming that the temperature of the gas in the initial state is equal to the virial temperature of the dark matter halo.
Using this threshold, the total energy injected in $5 \,\text{Myr}$ suffices for seven injection events within the simulated timespan. 

Fig.~\ref{fig:idealised_projection} shows a volume weighted projection of the gas density, temperature, absolute velocity and the energy dissipation weighted Mach number of the shocks present after $5\, \text{Myr}$. For the temperature and density, we average over the logarithm of the corresponding quantity, while we average over the absolute value of the velocity to highlight the maximum velocities involved in the projection. The projected maps show that the model is very efficient in diluting the central regions near the black hole. This means that the accretion rate estimated in this regime would decrease immediately by orders of magnitude, resulting in a very tight self-regulation. As the projections were made shortly after the seventh injection occurred, we reach gas flows with very high velocities in the injection region, which slow down after leaving this immediate vicinity of the black hole.

Fig.~\ref{fig:E_time_periodic} shows the time evolution of the thermal and kinetic energy after individual injection events, as well as the average evolution with time. 
We can detect an injection event simply by a jump in total energy of the system owing to very frequent simulation outputs. The first output after this jump defines the zero-point in Fig.~\ref{fig:E_time_periodic}, which means that the thermal energy increases and the kinetic energy decreases subsequently. The initial energy injection is purely kinetic; however within $\sim 0.5 \, \text{Myr}$ about half is dissipated into thermal energy, mostly via shock dissipation. As the direction of the momentum kicks change after every injection event, the AGN model does not build up a coherent gas flow that could reach several tens to a hundred $\text{kpc}$.

As a further analysis tool, we use the shock finder described in \citet{2015MNRAS.446.3992S} to detect shocks and calculate their Mach number $\mathcal{M}$ and energy dissipation rate for each snapshot. The bottom right panel in Fig.~\ref{fig:idealised_projection} shows the energy dissipation weighted Mach number projection, which excludes all cells that do not belong to a shock, and Fig.~\ref{fig:M_Statistic_rho} shows the corresponding energy dissipation as a function of Mach number. For a hot, dilute gas, the shock Mach number mostly remains below $10$, while the shock strength increases with higher densities and correspondingly lower temperatures and sound speeds. Summing up all the energy that is dissipated in shocks, we are, up to a factor of order unity, able to reconstruct the feedback energy purely through post-processing analysis of the surrounding gas. This is possible for even moderate resolution, which opens up the possibility of studying the effects of shocks from AGNs on their surroundings even in cosmological simulations of galaxy formation, so that their behaviour can be compared to observations \citep[e.g.][]{2015ApJ...801...42D,2015MNRAS.448.2301M}. However, there are some technical challenges to this \citep[see][]{2016arXiv160407401S}, in particular concerning the treatment of the unresolved ISM structure in these simulations. Therefore, for now we restrict our analysis to the idealized setup and leave the study of shocks from AGN winds in cosmological simulations to future work.

The test simulations demonstrate that our kinetic feedback model can accelerate the gas in the injection region to several tens of thousands of $\text{km}\,\text{s}^{-1}$. This gas flow hits the surrounding medium and heats it via shock dissipation within time-scales of a $\text{Myr}$. A fraction of the energy will remain kinetic and ultimately decay via turbulent dissipation. This behaviour of the feedback injection is well converged, showing that our deposition of energy is not subject to significant numerical limitations on marginally resolved scales.

\section{Cosmological simulations}
\label{sec:CosmoSims}

Idealized test simulations such as those above cannot address the dynamics of self-regulated black hole growth. As this can only be meaningfully studied in calculations that follow cosmic structure formation and that also account for star formation, we now move on and examine the impact of our new model in hydrodynamical simulations of galaxy formation. This requires the full black hole model as described in Section~\ref{sec:Model}, including the seeding of SMBHs and the estimate of their accretion rates. 
Also, because gas cooling and heating, star formation, stellar evolution and feedback, as well as the chemical enrichment of the interstellar medium, are all crucial ingredients of galaxy formation, we account for these processes using the respective models described in \citet{2013MNRAS.436.3031V}. These are modified and extended as follows: 
\begin{itemize}
\item We use isotropic winds from star formation with $10 \%$ of the energy injected thermally \citep{2014MNRAS.437.1750M}, instead of purely kinetically with a bipolar orientation as in Illustris.
 \item We slightly adjust the scaling of the stellar wind model with redshift, metallicity and halo mass.
 \item Updated chemical yields and an improved metal advection algorithm are used, which has however negligible influence on the results discussed here. 
 \item Ideal magnetohydrodynamics is included based on a Powell cleaning scheme \citep{2011MNRAS.418.1392P,2013MNRAS.432..176P}.
 \item An improved gradient estimator and time integration scheme for the hydrodynamics is used \citep{2016MNRAS.455.1134P}, which improves the accuracy of the {\small AREPO} code.
\end{itemize}

We briefly summarize the changes due to the modifications in the stellar wind model (i.e.~the first two items) in Appendix~\ref{app:wind}, as the interplay between stellar and AGN feedback affects the overall galaxy population \citep{2013MNRAS.428.2966P} as well as black hole growth rates \citep{2015MNRAS.452.1502D}. The other modifications have a minor effect on the quantities examined in this work. We therefore focus this study on the black hole model and its parameters, and illustrate the relevance of the feedback efficiency, the accretion rate estimate and the black hole seeding model for the formation and evolution of galaxies as a function of their mass. A more detailed analysis of the other changes will be subject of a forthcoming paper (Pillepich et al., in preparation).
  
\subsection{The simulations}

We run a number of cosmological simulations of a periodic box with a side length of $30\,h^{-1}$ Mpc. As the large-scale modes of the matter power spectrum cannot be sampled in this comparatively small volume, our simulation does not contain structures as massive as the largest galaxy clusters observed in our Universe. However, we still follow the formation of $13$ objects more massive than $10^{13}\,\text{M}_\odot$ and more than $100$ haloes in the mass range between $10^{12}\,\text{M}_\odot$ and $10^{13}\,\text{M}_\odot$. This makes the simulations well-suited for testing the AGN model and for studying its impact on the formation and evolution of massive galaxies at the resolution of the Illustris simulation.

We adopt the cosmological parameters from the \textit{Planck} intermediate results \citep{2015arXiv150201589P}, $\Omega_\text{M} = 0.3089$, $\Omega_\Lambda = 0.6911$, $\Omega_\text{b} = 0.0486$, $h=0.6774$ and $\sigma_8 = 0.8159$ and use an \citet{1998ApJ...496..605E} matter power spectrum to produce initial conditions at redshift $z=127$. The initial conditions contain $384^3$ dark matter particles and the same number of gas cells at our default resolution. This implies an average gas cell and dark matter particle mass of $6.7\times 10^6\,h^{-1}\,\text{M}_\odot$ and $3.4\times 10^7\,h^{-1}\,\text{M}_\odot$, respectively, which is similar to the intermediate resolution Illustris simulation \citep[Illustris-2 in][]{2014MNRAS.444.1518V}. The corresponding softening length is $2$ comoving kpc with a maximum value of $1$ proper kpc for dark matter and stars. The softening for the gas cells depends on their volume and has a minimum of $0.25$ comoving kpc. The moderate number of simulation particles allows us to study the effect of each parameter of the black hole model individually, but it also comes with a severe drawback: At this resolution, the star formation rate predicted by the employed \citet{2003MNRAS.339..289S} model is not fully converged for haloes below $10^{12.5}\,\text{M}_\odot$ \citep{2014MNRAS.444..237P,2015MNRAS.452..575S}, as also shown in Appendix~\ref{app:convergence}. This entails important limitations, especially with regard to the comparison to observations.

To get a better idea of the behaviour at the low mass end of AGN host galaxies, we run an additional simulation with $2\times 768^3$ particles and cells and the same box size. In this run, all softening lengths are reduced by a factor of $2$ compared to the fiducial setup. The implied resolution corresponds to the Illustris-1 high resolution run. Additionally, we run a low-resolution test with $2\times 192^3$ particles and the same side length of the simulation box. The softening is increased by a factor of $2$ compared to the fiducial run. For each of the different resolutions, we also computed a dark matter only version to quantify the role of baryonic physics on the halo mass function.

We have also carried out a suite of simulations with $2\times 384^3$ particles in which the parameters of the black hole model were systematically varied by a factor of $4$ each in the direction that seemed most interesting. Table~\ref{table:sims} gives an overview of these simulations and their parameters. The set of simulations also includes one simulation in which the black holes are always in the quasar-mode, independent of their Eddington rate (labeled `\texttt{no kin.}').  Using identical initial conditions in our simulations allows a halo-by-halo comparison of all galaxy properties, facilitating a clean comparison of globally averaged properties and an interpretation of small changes in a meaningful way. We do so by matching the friend-of-friends groups in the different simulations in position space, followed by a verification that they are indeed the same structures by ensuring that they have at least half of their dark matter particles in common.\footnote{This is done by checking their particle IDs, which are unique identifiers set in the initial conditions.} We discard the few percent of haloes that could not be matched by these criteria and ignore them for the analysis; they are for the most part borderline cases where the friend-of-friends algorithm links two haloes across a feeble particle bridge in one simulation but not in the other. In addition to the matching of haloes across simulations at identical redshifts, we also match haloes of a given simulation at different times. We define the progenitor as the halo in the previous snapshot that contributes the most dark matter particles to a given halo, which allows us to study the evolution of individual haloes.

We base a substantial part of our analysis on a comparison of the same haloes in different simulations, thereby avoiding uncertainties due to the absolute halo abundance, which is affected significantly by box size and resolution effects. However, a detailed comparison to observational data requires a larger simulated volume and higher resolution. Achieving both at the same time is a computational challenge and is clearly beyond the scope of this paper. Simulations that reach this statistical power will be presented in future work.

\begin{table*}
\begin{tabular}{llllllllllll}
\hline
\noalign{\vskip 0.5mm}
Name                     & $n_\text{particles}$      & $m_\mathrm{DM}$,$m_\mathrm{baryon}$      & $\epsilon_\text{f,kin}$ & $\epsilon_r$ &  $\chi_{0}$ & $\beta$ & $M_\text{seed}$ & $M_\text{FOF}$ & $n_\text{ngb}$ & $f_\text{re}$\\
\noalign{\vskip 0.5mm}
                         & \small{in initial conditions}  & \small{[$10^5\,\text{M}_\odot\,h^{-1}$]} &  &  &  &   & \small{[$\text{M}_\odot\,h^{-1}$]} & \small{[$\text{M}_\odot\,h^{-1}$]} &  &\\
\noalign{\vskip 0.5mm}
\hline
\hline
Fid             & $2\times 384^3$    & $340,\,67$    & $0.2$    & $0.2$    & $0.002$    &  $2.0$    & $8\times10^5$    & $5\times 10^{10}$ & $128$ & $20$\\
\noalign{\vskip 0.5mm}
No kin.             & $2\times 384^3$    & $340,\,67$    & $0.2$    & $0.2$    & $\boldsymbol{0.000}$    &  $2.0$    & $8\times10^5$    & $5\times 10^{10}$ & $128$ & $20$\\
\noalign{\vskip 0.5mm}
High res    & $\boldsymbol{2\times 768^3}$    & $\boldsymbol{42.5,\,8.4}$    & $0.2$    & $0.2$    & $0.002$    & $2.0$    & $8\times10^5$    & $5\times 10^{10}$ & $\boldsymbol{256}$ & $20$\\
\noalign{\vskip 0.5mm}
Low res    & $\boldsymbol{2\times 192^3}$    & $\boldsymbol{2720,\,536}$    & $0.2$    & $0.2$    & $0.002$    & $2.0$    & $8\times10^5$    & $5\times 10^{10}$ & $\boldsymbol{64}$ & 20\\
\hline
\noalign{\vskip 0.5mm}
Low $\epsilon_{f}$    & $2\times 384^3$    & $340,\,67$    & $\boldsymbol{0.05}$    & $0.2$    & $0.002$    & $2.0$    & $8\times10^5$ & $5\times 10^{10}$ & $128$ & $20$\\
\noalign{\vskip 0.5mm}
Low $\epsilon_r$    & $2\times 384^3$    & $340,\,67$    & $\boldsymbol{0.05}$    & $\boldsymbol{0.05}$    & $0.002$    & $2.0$    & $8\times10^5$    & $5\times 10^{10}$ & $128$ & $20$\\
\noalign{\vskip 0.5mm}
High $\chi$    & $2\times 384^3$    & $340,\,67$    & $0.2$    & $0.2$    & $\boldsymbol{0.008}$    & $2.0$    & $8\times10^5$    & $5\times 10^{10}$ & $128$ & $20$\\
\noalign{\vskip 0.5mm}
Low $\beta$    & $2\times 384^3$    & $340,\,67$    & $0.2$    & $0.2$    & $0.002$    & $\boldsymbol{0.5}$    & $8\times10^5$    & $5\times 10^{10}$ & $128$ & $20$\\
\noalign{\vskip 0.5mm}
Low $M_\text{seed}$    & $2\times 384^3$    & $340,\,67$    & $0.2$    & $0.2$    & $0.002$    & $2.0$    & $\boldsymbol{2\times10^5}$    & $5\times 10^{10}$ & $128$ & $20$ \\
\noalign{\vskip 0.5mm}
High $M_\text{FOF}$    & $2\times 384^3$    & $340,\,67$    & $0.2$    & $0.2$    & $0.002$    &  $2.0$    & $8\times10^5$   & $\boldsymbol{2\times 10^{11}}$ & $128$ & $20$\\
\noalign{\vskip 0.5mm}
High $n_\text{ngb}$    & $2\times 384^3$    & $340,\,67$    & $0.2$    & $0.2$    & $0.002$    & $2.0$    & $8\times10^5$  & $5\times 10^{10}$ & $\boldsymbol{512}$ & $20$\\
\noalign{\vskip 0.5mm}
Low $f_\text{re}$    & $2\times 384^3$    & $340,\,67$    & $0.2$    & $0.2$    & $0.002$    & $2.0$    & $8 \times 10^5$   & $5\times 10^{10}$ & $128$ & $\boldsymbol{5}$\\
\noalign{\vskip 0.5mm}

\hline

\end{tabular}
\caption{Overview of our primary simulations. All simulations, except for high res and low res are started from the same initial conditions. The only differences between the simulations and the fiducial case are the parameter values marked in bold. All simulations have a comoving volume of $(30\,h^{-1} \text{Mpc})^3$.}
\label{table:sims}
\end{table*}


\subsection{Galaxy properties}

\begin{figure}
 \centering
 \resizebox{8.5cm}{!}{\includegraphics{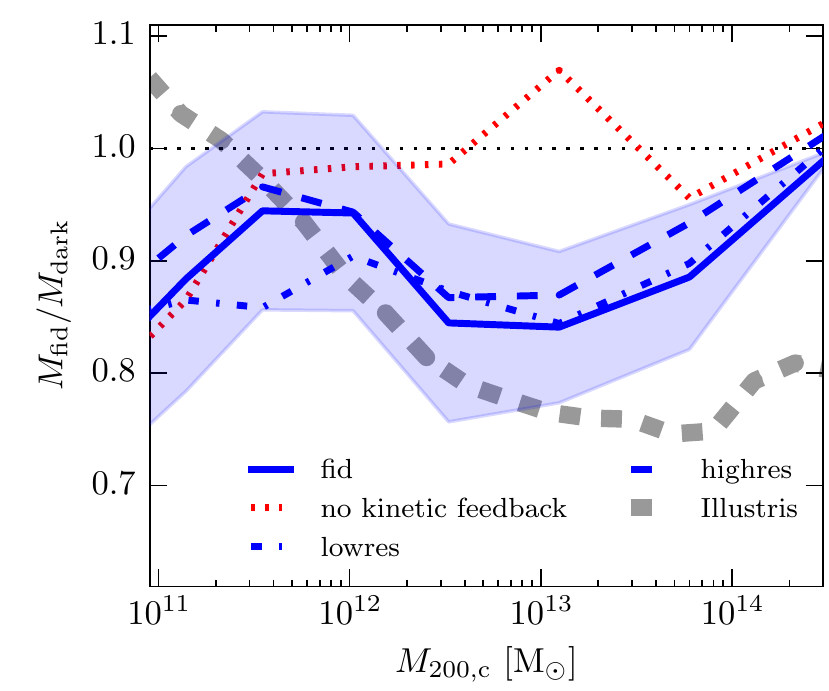}}
 \caption{Average ratio of the mass of haloes in the full physics simulation to the mass of  the corresponding halo in the dark matter only run, as a function of halo mass. The shaded region indicates the $1\sigma$ scatter
   of the results of our default simulation. The dashed grey line represents the result for the Illustris simulation \citep{2014MNRAS.444.1518V}. }
 \label{fig:massfraction}
\end{figure}

\subsubsection{Halo masses}

We start by looking at the overall baryonic effect on the masses of individual haloes. Fig.~\ref{fig:massfraction} shows the mass of the haloes in our fiducial baryonic simulation in units of the mass of the corresponding halo in the dark matter only simulation, as a function of $M_\text{200,c}$. The average mass fraction does not exceed unity for any halo mass, unlike in \citet{2014MNRAS.444.1518V} for haloes of $M_{200,c}\sim 10^{11}\, \text{M}_\odot$. However, also in our simulations, the mass of some individual haloes can scatter above the mass of their dark matter only counterparts. The masses of haloes with $M_{200,c} < 10^{11}\, \text{M}_\odot$ are suppressed more than those of $10^{12} \,\text{M}_\odot$ haloes, independent of the black hole feedback implementation. This is hence presumably caused by stellar feedback. A mild decline in the halo masses relative to the dark matter only run occurs for haloes more massive than $10^{12}\, \text{M}_\odot$. This drop can be clearly associated with the kinetic AGN feedback, as it is not present in the simulation without this mode. However, it is not as pronounced as in \citet{2014MNRAS.444.1518V}, which confirms that our feedback implementation is not as violent. The upturn at $10^{14}\, \text{M}_\odot$ indicates a return to the universal baryon fraction for the most massive haloes but is based only on very few haloes. Better statistics will be needed to reliably establish the behaviour at these mass scales.


\subsubsection{Black holes}

\begin{figure}
 \centering
 \includegraphics{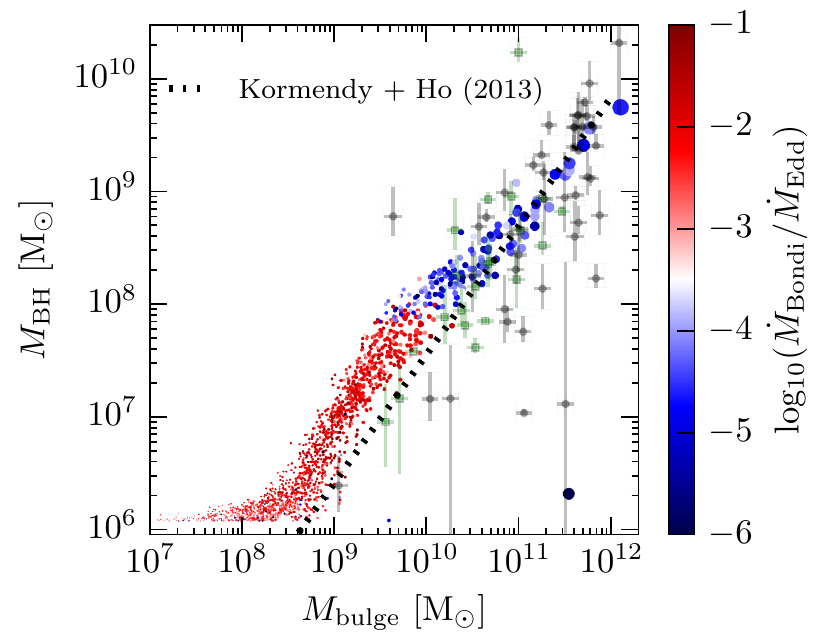}
 \includegraphics{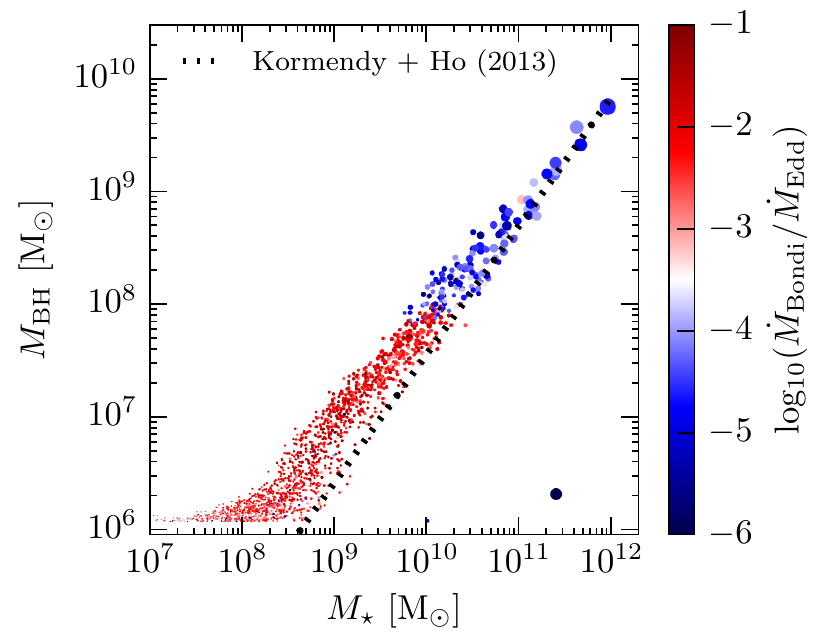}
 \caption{Black hole mass as a function of bulge mass {(upper plot) and stellar mass within twice the half mass radius (lower plot)} for central galaxies in
   the high-resolution simulation. The size of the symbols is scaled with bulge mass for better visibility, and the assigned colour scale encodes the Eddington ratio. The dotted line is the fit to
   observational data. The symbols with error bars are observed
   ellipticals (black), and spirals or S0 galaxies with normal bulges
   (green), taken from \citet{2013ARA&A..51..511K}. The
   bulge mass is estimated as twice the mass of the counterrotating
   fraction of stars within $0.1\,R_{200,c}$. We note that this might slightly
   underestimate the bulge mass in the case of rotating bulges.}
 \label{fig:Mbh_Mstar}
\end{figure}
\begin{figure}
 \includegraphics{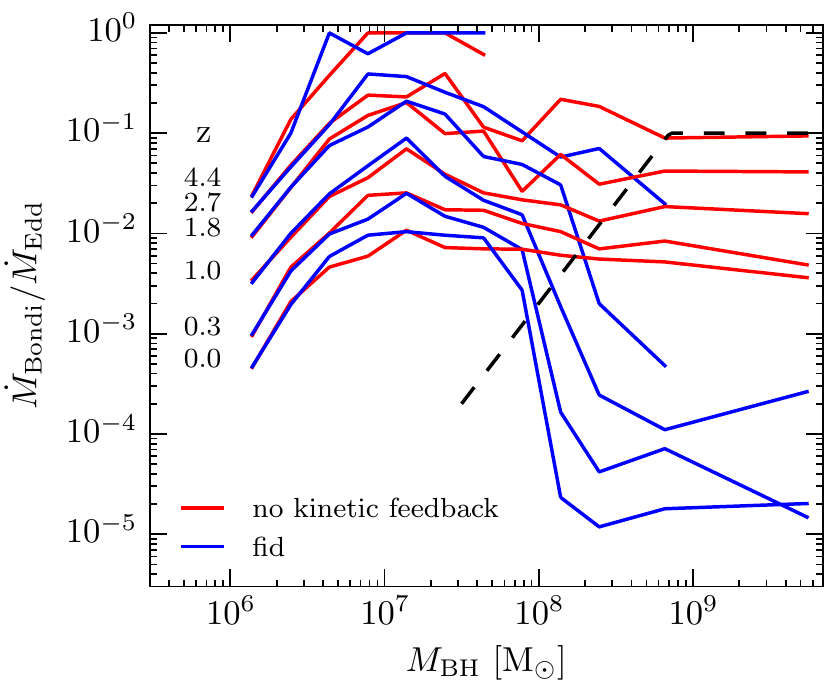}
 \includegraphics{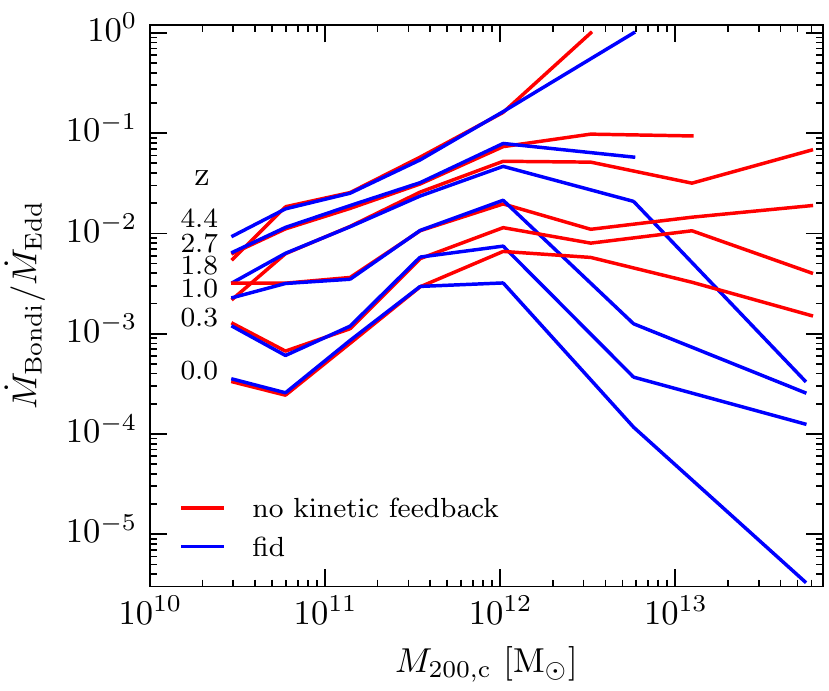}
 \caption{Median Eddington ratio as a function of black hole mass
   (top) and halo mass (bottom). The different lines show
   different redshifts. The dashed line in the top plot indicates our imposed
   transition point between low and high accretion states.}
 \label{fig:fEdd_Mass}
\end{figure}

\begin{figure}
 \center
 \resizebox{8.5cm}{!}{\includegraphics{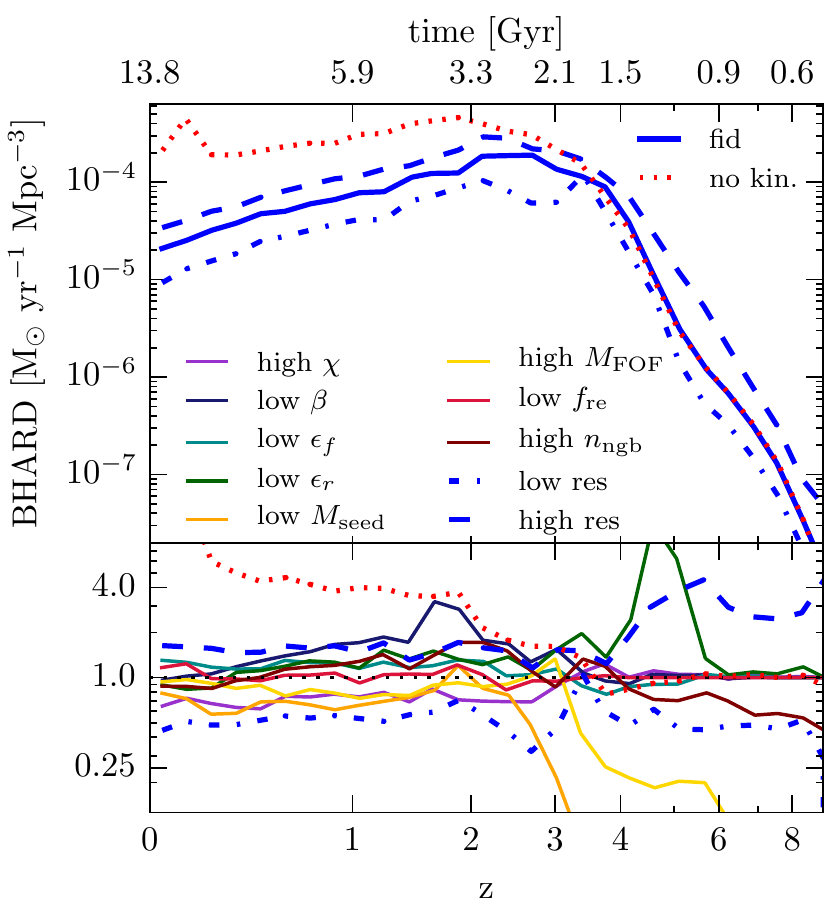}}
 \caption{BHARD as a function of redshift
   for different runs. The lower panel shows the ratio relative
   to the fiducial run.}
 \label{fig:BHARD_Z}
\end{figure}

As mentioned in Section~\ref{subsec:model_seeding}, a black hole with mass $M_\text{seed}$ is placed into a halo whenever the on-the-fly friend of friends halo finder identifies a structure that is more massive than a threshold mass $M_\text{FOF}$ and does not yet contain a black hole. At seeding, the surrounding gas is usually dilute and the black hole accretes at low rates with a long Bondi growth time-scale. This means that the growth is slower than the growth in stellar mass and therefore, the corresponding galaxy evolves horizontally in the $M_\text{BH}-M_\text{bulge}$ diagram (Fig.~\ref{fig:Mbh_Mstar}{, upper plot}). After some time, enough gas piles up around the black hole and produces higher accretion rates, allowing the black hole to eventually grow more rapidly, aided also by the runaway character of Bondi growth due to its $\dot M_{\rm BH}\propto M_{\rm BH}^2$ scaling. Consequently, the slope in the $M_\text{BH}-M_\text{bulge}$ diagram steepens. This second phase continues until the feedback injection of the black hole into its surroundings becomes significant, at which point the black hole gas supply becomes self-regulated. In this final stage, the black holes grow less rapidly and are mostly in the low accretion state again.

However, the slight change in slope in the $M_\text{BH}-M_\text{bulge}$ relation at $M_\text{bulge}\approx 10^{10}\,\text{M}_\odot$ is not due to the change of accretion mode, but rather due to the bulge-to-disc decomposition. We define the bulge mass as twice the stellar mass of the counterrotating star particles within $0.1\,R_{200,c}$. Galaxies with $M_\text{bulge}\approx 10^{10}\,\text{M}_\odot$ have a large fraction of corotating stars (i.e. a disc) and correspondingly our estimate of the bulge mass is reduced, which shifts the corresponding points to the left in {the upper plot of} Fig.~\ref{fig:Mbh_Mstar}. In the $M_{\rm BH}-{M}_{*}$ plot {(Fig \ref{fig:Mbh_Mstar}, lower plot)}, with ${M}_{*}$ being the mass of all the stars within twice the stellar half-mass radius, such a change in slope does not show up.

Generally speaking, our systems with $M_* < 10^{10.5}\, \text{M}_\odot$ tend to have slightly overly massive black holes compared to the observed relation, which indicates too early growth of the black holes, possibly caused by the increased seeding mass we use compared to earlier work \citep{2015MNRAS.452..575S}. However, \citet{2016arXiv160201941V} showed that the shape and scatter of the low-mass end changes significantly for different ways to measure $M_\text{bulge}$. Considering this effect and the observational uncertainties, the discrepancy is not particularly worrisome. For high-mass systems, we follow the observed relation more closely, seemingly with little scatter. We leave a detailed analysis of the high mass end to future work as it requires a larger sample of black holes.

The black hole population has a clear change in accretion rate at black hole masses of around $10^8\,\text{M}_\odot$ (colour coded in Fig.~\ref{fig:Mbh_Mstar}). Fig.~\ref{fig:fEdd_Mass} shows the accretion rate in units of $\dot{M}_\text{Edd}$ as a function of black hole mass for different redshifts. For comparison, we also show the Eddington factors in the run without kinetic feedback. One clear trend is the drop of the Eddington ratio with redshift, which is consistently present over the complete range of black hole masses. This is expected, as the black hole accretion rate density (BHARD) (Fig.~\ref{fig:BHARD_Z}) in $\text{M}_\odot\,\text{yr}^{-1} \text{Mpc}^{-3}$ decreases significantly towards low redshifts. The downturn towards the low black hole mass end shows the relatively slow initial growth of the black holes. This is partially due to a smaller amount of cold gas and partially due to the fact that in our implementation the Eddington ratio depends linearly on the black hole mass. A perhaps unexpected feature is that, without kinetic feedback, the Eddington factor does not significantly vary with black hole mass for black holes more massive than $10^7\,\text{M}_\odot$. This means that even the most massive black holes in galaxy group and cluster environments would have a good chance of accreting at high Eddington ratios, which is almost invariably also associated with high star formation rates.

This can be prevented by the kinetic feedback of the low accretion mode, which is more efficient than the feedback in the high accretion state. To ensure that a black hole and its surrounding gas transition to a self-regulated state with lower accretion rate, and to prevent newly seeded low-mass black holes from remaining in the low accretion state, we employ a black hole mass dependent quasar threshold $\chi$, shown as the dashed line in the top panel of Fig.~\ref{fig:fEdd_Mass}. Once a black hole transitions to the kinetic feedback mode, its Eddington factor drops significantly as a consequence of the stronger feedback, making it likely to remain in this regime for an extended period of time.  Note that the deviation of the median curve starts slightly to the left of the dashed line due to the scatter in the black hole properties, allowing some lower mass black holes to make the transition earlier than the mean.

The drop in Eddington ratio for high-mass systems also has an effect on the overall BHARD, shown in Fig.~\ref{fig:BHARD_Z}. In fact, for the simulations without kinetic feedback, there is no significant drop of the BHARD towards lower redshifts. There is an increased BHARD with higher resolution, which is related to the fact that the region in which the accretion rate is estimated is intentionally reduced with increasing resolution, which leads to systematically higher density estimates. Especially at early times, this leads to earlier and thus faster accretion, and consequently more massive black holes.


\subsubsection{Stellar component}

\begin{figure}
 \centering
  \resizebox{8.5cm}{!}{\includegraphics{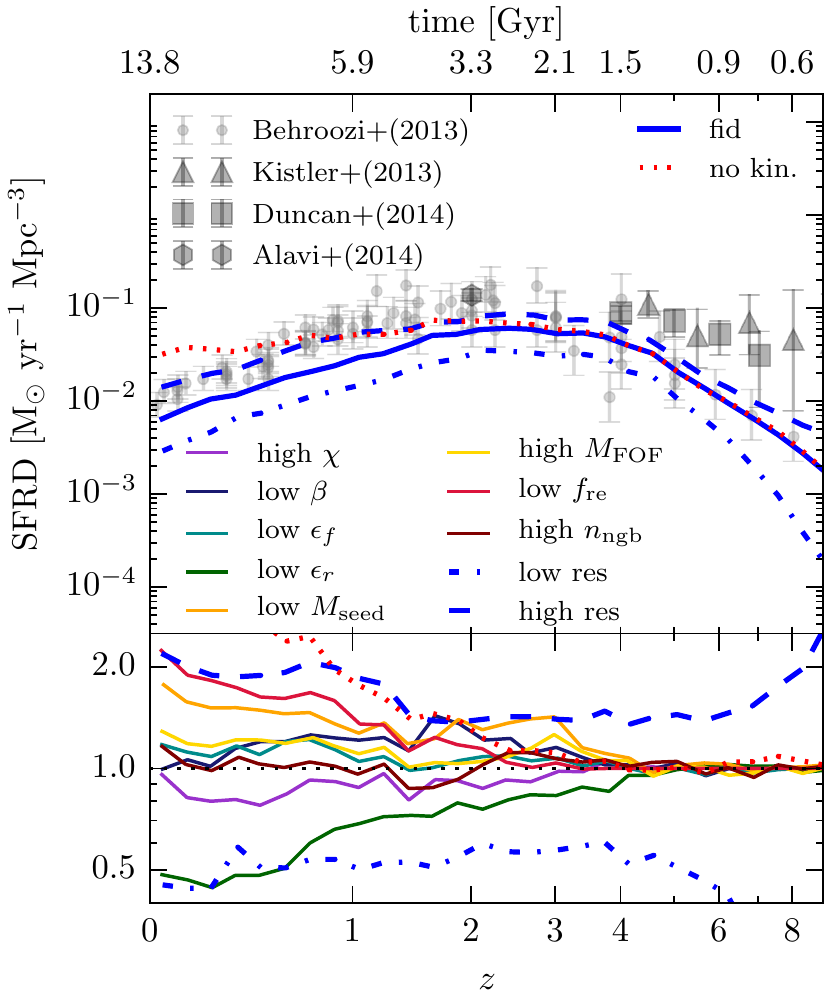}}
  \caption{SFRD as a function of
    redshift. The grey dots{, triangles, squares and hexagons} are observational data from
    \citet{2013ApJ...762L..31B,2013ApJ...770...57B}{, \citet{2013arXiv1305.1630K}, \citet{2014MNRAS.444.2960D} and \citet{2014ApJ...780..143A}, respectively}. The lower panel
    shows the ratio of the SFRD relative to the fiducial run.}
 \label{fig:SFRD-Z}
\end{figure}

\begin{figure}
 \centering
 \includegraphics{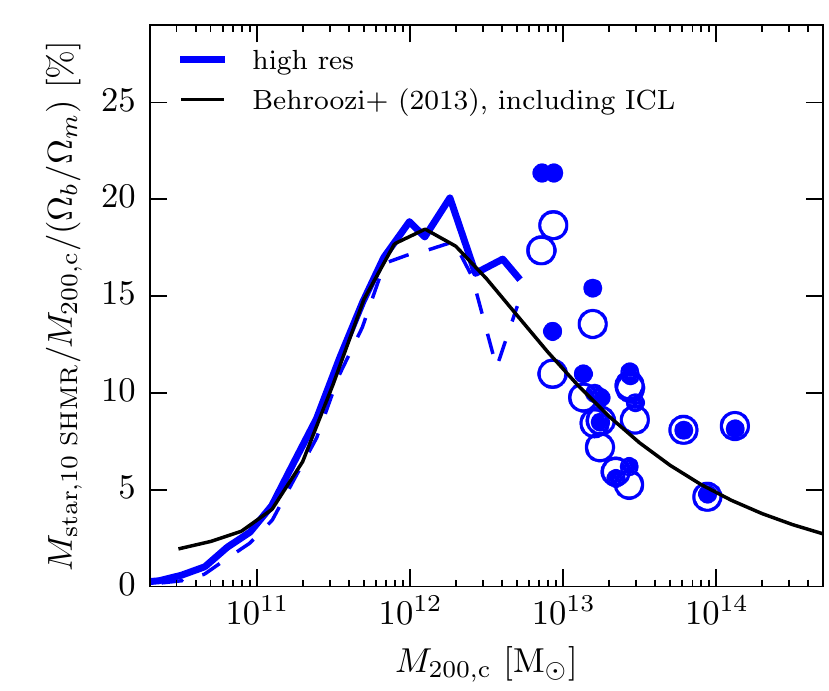}
 \caption{Stellar mass fraction as a function of halo mass for the
   high-resolution simulation. The stellar mass is calculated as the
   mass of all star particles within $10$ stellar half-mass radii that
   do not belong to a subhalo. The black line is the corresponding fit
   to observations from
   \citet{2013ApJ...762L..31B,2013ApJ...770...57B} including the
   intracluster light. We use the simulation values of
   $\Omega_\text{b}$ and $\Omega_\text{m}$ for both simulation and
   literature data. The dashed line and open circles correspond to the
   same simulation data, but the halo mass $M_\text{200,c}$ is taken
   from the corresponding halo in the dark matter only simulation
   \citep[see][for a discussion]{2013ApJ...766...56M}.}
 \label{fig:fStarICL_Mhalo}
\end{figure}

\begin{figure}
 \centering
 \includegraphics{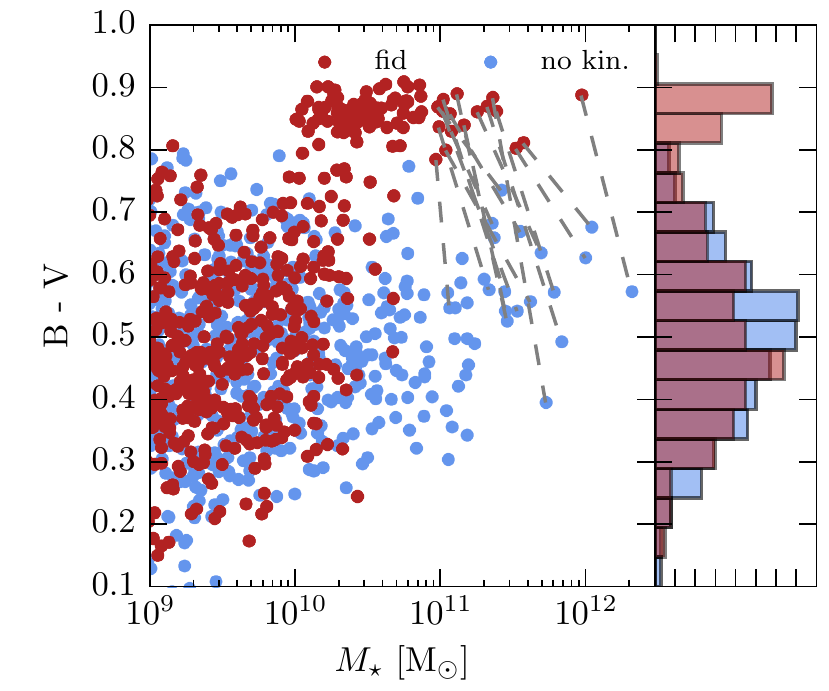}

 \includegraphics{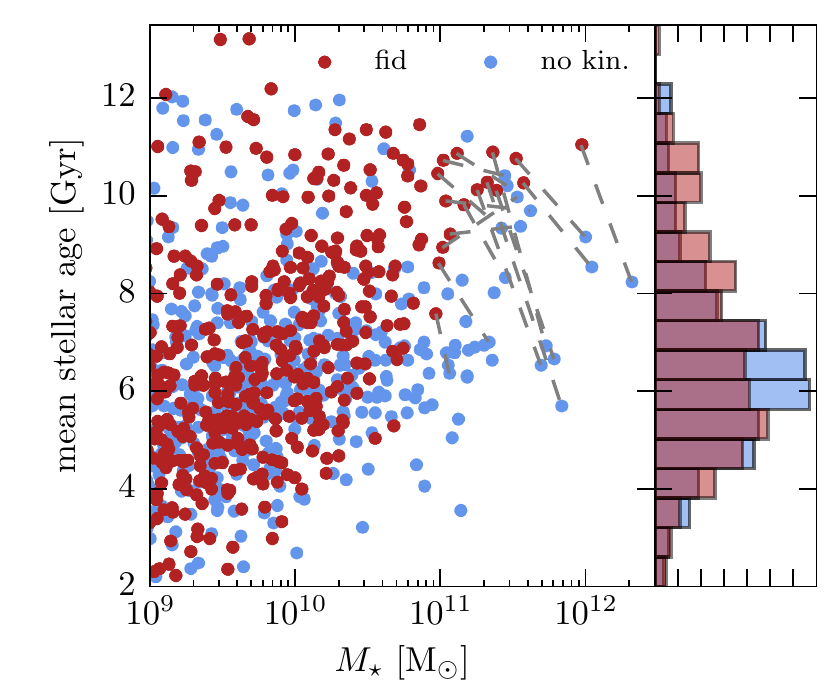}
 
 \caption{{Top panel:} $B$-$V$ colours as a function of stellar mass within twice the
   stellar half mass radius. The red dots are from the fiducial
   simulations. The blue dots represent the haloes in the run without
   kinetic feedback. For the most massive systems, the dashed lines
   link the same haloes in the two different runs to emphasize the
   effect of the kinetic feedback mode on a halo-by-halo basis. The histogram on the side clearly shows the emergence of a red (larger $B$-$V$ values) population of galaxies due to the kinetic feedback model. { Bottom panel:} mass-weighted stellar age as a function of stellar mass
   within twice the stellar half mass radius. Note that the choice of colours differs from the other figures.}
 \label{fig:BminusVvsStellarmass}
\end{figure}

We now turn to the effect of black hole feedback on the stellar properties of galaxies.  Fig.~\ref{fig:SFRD-Z} shows the average star formation rate density (SFRD) as a function of redshift. At redshifts lower than $z=3$ the fiducial simulation differs from the observed SFRD by about $0.3$ dex. For our higher resolution simulation (dashed line), the SFRD is in better agreement with the observations. At low redshifts, the contribution from Milky Way-sized galaxies dominates, which indicates a relatively poor convergence in their star formation rates. In Appendix~\ref{app:convergence}, we discuss this in more detail.

From an AGN-feedback point of view, galaxy formation can be divided into three epochs.  At redshift $z> 5$, there are either no or only slowly accreting black holes with no significant impact on the host galaxy. During this stage, only stellar feedback regulates the star-formation rate \citep[e.g.][their fig. 15]{2013MNRAS.436.3031V}. Correspondingly, changes in the black hole parameters have no effect on the SFRD. After this initial phase, the black holes enter the high accretion regime and grow quickly, releasing a considerable amount of thermal feedback energy that suppresses star formation, in particular in galaxies with a final halo mass $>10^{12}\,{\rm M}_\odot$.  At late times, from redshift $z=2$ to the present day, the black holes switch to the low-accretion regime again, remaining in a self-regulated state in which both the stellar and AGN feedback balance cooling. The relative importance of AGN over stellar feedback depends on halo mass. While stellar feedback dominates in haloes up to the size of the Milky Way, more massive haloes are mainly regulated through AGN feedback.

The stellar mass fraction as a function of halo mass (Fig.~\ref{fig:fStarICL_Mhalo}) clearly shows the decrease in star formation efficiency with halo mass at the massive end, in good agreement with observations. We find it useful to compare the stellar mass including the diffuse intra-cluster light in the high mass end in observations and theory. In this way, we are less sensitive to the choice of the aperture within which stellar masses are estimated.  {At the high halo-mass end, the data shown here are in reasonable agreement with \citet{2014arXiv1401.7329K} who pointed out the importance of outer stellar profiles in high-mass haloes in this type of analysis}. Additionally, we plot both the halo mass from the full physics simulation as well as the halo mass of the corresponding halo in the dark matter only run, where the latter, i.e. the dashed line and open circles, should be compared to the results from abundance-matching. Fig.~\ref{fig:massfraction} shows the ratio of these two masses as a function of halo mass. In particular at around $M_\text{200,c}\approx 10^{13} \, \text{M}_\odot$ it turns out to be crucial to take this effect into account.

One of the conjectured effects of AGN is that they can prevent the most massive galaxies from being blue and star forming, instead making them red and having an old stellar population. We use $B$-$V$ colour and mean stellar age (Fig.~\ref{fig:BminusVvsStellarmass}) as a measure for the efficiency of the feedback in the low accretion state to accomplish this.  To probe the relevance of the kinetic feedback mode for this, we compare our fiducial simulation with a simulation without the kinetic mode. The panels of Fig.~\ref{fig:BminusVvsStellarmass} clearly show the need for this efficient mode to get `red and dead' galaxies with old stellar populations on the massive end of the galaxy population.

\subsubsection{Gas component}

\begin{figure}
 \centering
 \includegraphics{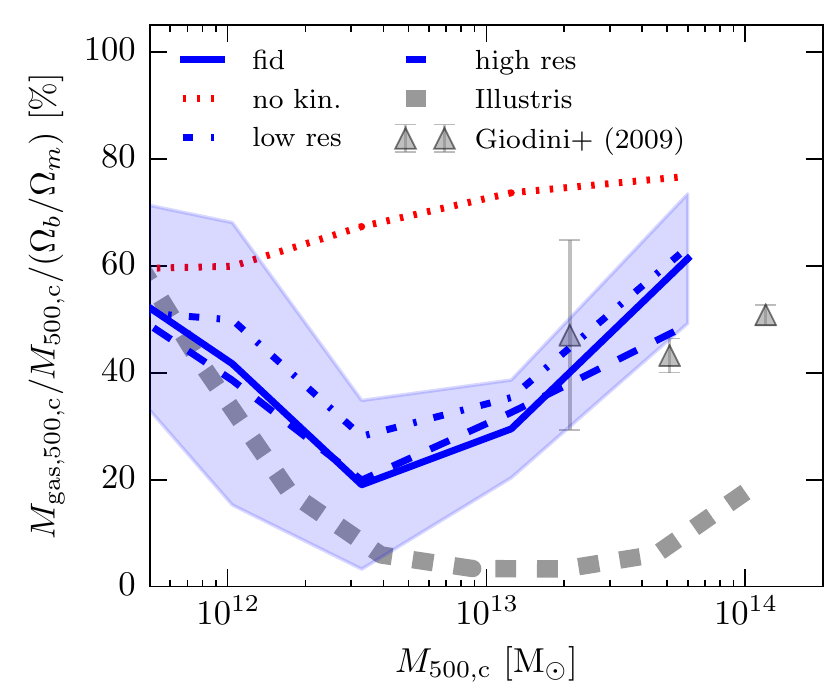}
 \caption{Gas mass fraction in $R_{500,c}$ as a function of halo mass
   $M_{500,c}$. The triangles show the binned data from
   \citet{2009ApJ...703..982G}. The dashed grey line represents the gas fractions in the Illustris simulation \citep{2014MNRAS.445..175G}. }
 \label{fig:gasmassfraction}
\end{figure}

\begin{figure}
  \centering
  \includegraphics{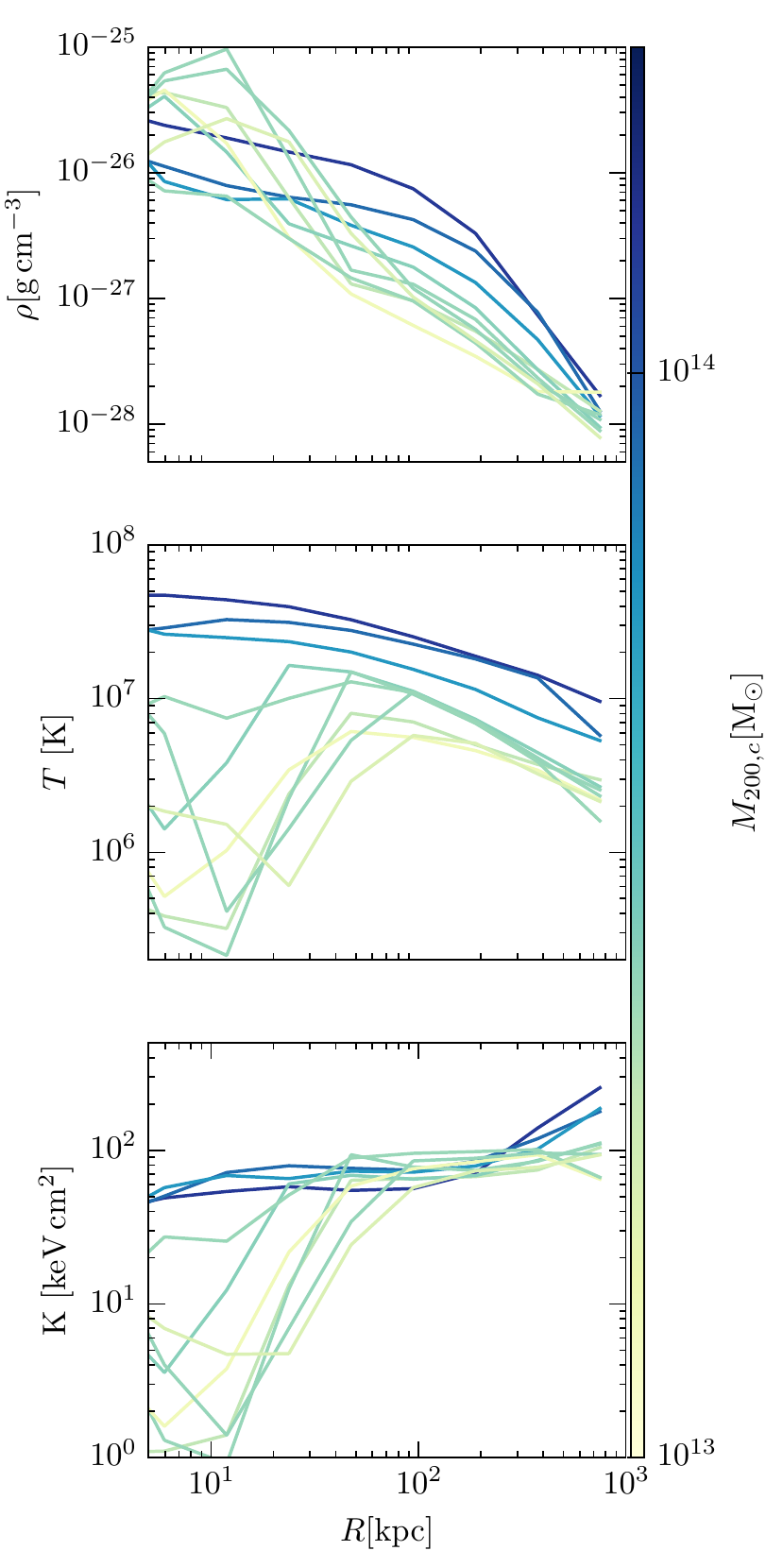}
  \caption{Mass-weighted density (top), temperature (middle) and
    entropic function (bottom) of the $10$ most massive haloes in the
    high-resolution simulation. The line colour encodes $M_{200,c}$ of
    the halo.}
  \label{fig:profiles}
\end{figure}

The low gas fraction of haloes around $10^{13}\,\text{M}_\odot$ has been identified as one of the main shortcomings of the Illustris simulation \citep{2014MNRAS.445..175G}. The gathering of substantial amounts of feedback energy invoked in the bubble model of Illustris, and its explosive release once enough energy is available, does prevent the feedback energy from being quickly lost due to cooling, but it also expels a significant fraction of gas from the inner halo. This resulted in a gas fraction which is factor of a few too low in systems where the feedback is most efficient.  In Fig.~\ref{fig:gasmassfraction}, we show the gas fractions within $R_{500,c}$ as a function of their mass $M_{500,c}$, obtained with our new kinetic feedback model. Reassuringly, it does not expel too much gas from the inner halo, but rather heats it via shocks and drives turbulence in the halo core, leading to an overall good agreement with observations.

To further investigate the effect of AGN feedback on the gas properties it is instructive to look at the radial profiles of the gas distribution. To this end, we use the high resolution simulation and plot the density, temperature and entropic function profiles in Fig.~\ref{fig:profiles}. For the most massive haloes with a mass around $10^{14}\, \text{M}_\odot$, the temperature profiles are almost flat in the centre and the central entropic function $K=~k_\text{B}\,T\,n^{-2/3}$ has a value of around $50 \,{\rm k}\text{eV}\,\text{cm}^2$. This confirms that the efficient quenching of star formation is not due to overly heating and diluting the central gas. For haloes less massive than $10^{13.5}\, \text{M}_\odot$, the density profiles are more centrally peaked and the temperatures in the centres are lower, which indicates that these haloes might have some residual star formation. As the volume is relatively small, our simulations do not contain massive galaxy clusters for which we could compare the thermodynamic profiles with observations of local galaxy clusters. Such simulations of galaxy clusters and how they are impacted by different AGN models are analysed in detail in forthcoming work (Popa et al., in preparation).


\section{Dependence on model parameters}
\label{sec:parameters}

\begin{figure*}
\includegraphics{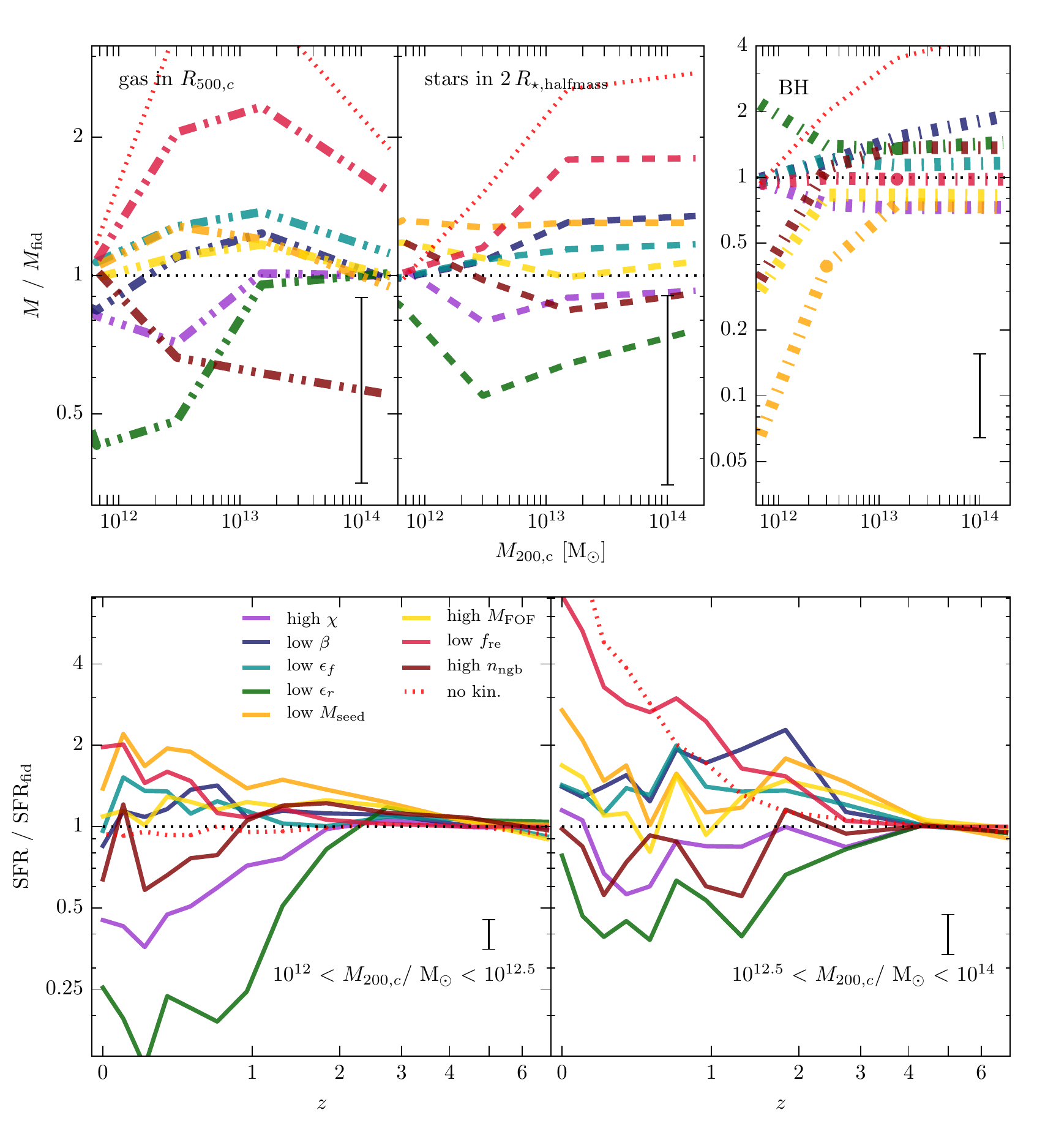}
\caption{{ Upper panels (from left to right):} gas mass in $R_{500,c}$,
  stellar mass in twice the stellar half-mass radius and black hole
  mass relative to their values in the fiducial run, compared on a
  halo-by-halo basis and binned as a function of the fiducial
  $M_{200,c}$. For clarity, we do not plot the individual scatter, but
  indicate with the black error bar the average scatter. { Lower panels:}
  average star formation rate of systems of different final mass (in
  the fiducial run) in units of their star formation rate in the
  fiducial simulation at a given redshift, on a halo-by-halo basis as
  a function of redshift. The error bar indicates the uncertainty of the
  mean.}
\label{fig:paramstudy}
\end{figure*}

We now investigate how robust the findings discussed in Section~\ref{sec:CosmoSims} are against changes in the parameters of our new black hole model. As we run identical initial conditions with several different parameter settings, we can compare their effects on a halo-by-halo basis.  Figure~\ref{fig:paramstudy} shows the relative changes in the gas-, stellar- and black hole masses binned with respect to halo mass, as well as the star formation rate as a function of redshift for systems with different halo mass at $z=0$. We shall first discuss the variations due to modifications of the efficiency parameters $\epsilon_f$ and $\epsilon_r$, and then consider the other parameters in turn.

\subsection{Global properties}

For all the investigated changes of model parameters, there is generally only a weak change in the late time accretion rate density (Fig.~\ref{fig:BHARD_Z}). At higher redshift, the seeding parameters have however a significant impact on the accretion and growth history of the black holes. In our tests, we lower the seed mass $M_\text{seed}$ or increase the halo mass $M_\text{FOF}$ at which black holes are seeded, which both delay black hole growth. Lowering the radiative efficiency $\epsilon_r$ leads to a higher Eddington accretion limit and therefore an increased accretion rate at $z=5$. As soon as the accretion rate is feedback regulated, it drops back to the fiducial rate. There is also an increase in the BHARD around $z=2$ for the simulation with low quasar threshold slope $\beta$. This can be explained by a delayed transition to kinetic feedback for the black hole population.

We now focus on the change in the SFRD for different parameter settings, shown in Fig.~\ref{fig:SFRD-Z}. The most significant variations occur for the simulations with modified values of $M_\text{seed}$, $\epsilon_r$ or $f_{\rm re}$. Changes in these parameters manifest themselves in the global SFRD after redshift $z=4$, and the effect increases at later times. However, it is also evident that a factor of $2$ change in spatial resolution has a stronger impact on the global SFRD than a factor of $4$ change in any of the black hole parameters.  The significant decline in SFRD for a lower $\epsilon_r$ can be explained by the increased accretion rate due to the lower feedback energy per accreted mass. The over-massive black holes in turn have a significantly larger impact on the star formation rate in the host galaxies. This will be further investigated in Section~\ref{subsec:efficiency}.

\subsection{Halo-by-halo comparison}

\subsubsection{Efficiency parameters}
\label{subsec:efficiency}

\paragraph*{}The feedback efficiency $\epsilon_{\rm f, kin}$ is defined as the fraction of accreted rest mass energy that appears kinetically in the kinetic wind mode. Lowering this efficiency by a factor of $4$ reduces the amount of feedback energy, which leads to an increase in the gas fraction and stellar mass, in particular in the high-mass systems. Having a higher gas fraction and star formation rate indicates that the central gas is denser and has lower temperatures than in the fiducial run. This leads to slightly increased black hole masses. Overall, the effect of this drastic change in kinetic feedback efficiency is rather small, which can be explained by the self-regulated nature of the feedback cycle.

\paragraph*{}The radiative efficiency $\epsilon_r$ determines how much of the accreted mass is converted to AGN luminosity in the quasar mode. To achieve a reduction of feedback energy for all black holes independent of accretion mode, we also reduce $\epsilon_\text{f,kin}$ in the corresponding test. Lowering the overall feedback efficiency by a factor of $4$ also increases the Eddington limit by the same factor, which leads to a significantly faster growth of the black holes (bump at $z=5$ in Fig.~\ref{fig:BHARD_Z}), but leaves the injected feedback energy for a black hole accreting at that limit constant, given the same black hole mass. As the black holes accrete more, they become more massive and have therefore a more significant impact on their surroundings. This increases the quenching in all systems and expels gas from Milky Way-sized galaxies. As the mass of the black hole increases, the threshold for them to be in the kinetic feedback mode also increases; i.e.~more energy is injected in this mode. This might be an additional amplifying factor for the low gas fractions and the efficient quenching of these systems. The fact that a lowering of the quasar-threshold has similar effects (see below) indicates that this is indeed the case.


\subsubsection{Accretion rate dependences}
\label{subsec:accretion}

\paragraph*{}The quasar threshold $\chi$ determines whether a black hole is associated with the low or high accretion rate state. This means that in our test simulation (higher $\chi$) the Bondi accretion estimate is four times higher when it transitions from the high accretion state to the low accretion state. This seems to have no effect on the initial growth of the black holes, which indicates that the black holes easily exceed the threshold at early times and accrete most of their mass in the high accretion state. At lower redshift, however, the average Eddington rates decrease to a level where the increase of the threshold by a factor of four matters. As our model involves a mass dependence of $\chi$, the higher Eddington rate threshold can also be interpreted as a lowering of the black hole mass for which, at a fixed Eddington factor, a black hole transitions between thermal and kinetic feedback modes. As the black hole mass correlates with halo mass, this means that the mass scale at which black holes are predominantly in the kinetic mode is effectively shifted to lower masses. And because the kinetic feedback mode is comparatively more efficient at quenching a halo, this explains the dip in gas and stellar mass at around $3\times 10^{12}\,\text{M}_\odot$. The lowered black hole masses above this mass scale can be explained by the fact that the black holes only grow significantly in the quasar mode. When the kinetic mode is switched on earlier, the black holes end up systematically less massive, provided they reach the transition threshold in the first place.

\paragraph*{}The slope $\beta$ influences the adopted scaling of the Eddington ratio with black hole mass for setting the transition between quasar- and kinetic mode. Reducing it by a factor of four as done in our test means that low mass black holes (below $10^8\, \text{M}_\odot$) will be found more often in the kinetic mode, and higher mass black holes more often in the quasar mode, compared to our default run. This explains the relative increase in black hole mass towards high mass haloes, keeping in mind that black holes predominantly grow in the quasar mode. Knowing that the quasar mode is less efficient at quenching, this also explains the larger stellar masses for high-mass systems as well as the fact that the additional stars form mainly at higher redshift, at the time when the haloes are delayed in switching to the kinetic mode.


\subsubsection{Black hole seeding parameters}
\label{subsec:seeding}

\paragraph*{}The black halo seed mass $M_\text{seed}$ is the initial mass given to the black holes when they are inserted in newly emerging haloes. If the black holes begin their evolution with smaller masses as in our test, their growth time-scale is considerably longer, because the Bondi accretion rate scales as $\dot{M}_\text{Bondi}\propto M_\text{BH}^2$. This means that the black holes grow significantly later, as more gas needs to accumulate in the halo centres to start a rapid growth. This delayed growth implies that the black holes in $10^{12}\,\text{M}_\odot$ systems have not yet ended their rapid accretion phase at $z=0$. This explains why the mean black hole mass is an order of magnitude below the mass in the fiducial run. For the more massive haloes, the black holes formed earlier and already had enough time to catch up; however, they are still about $25 \%$ less massive. The delayed growth also implies that there is less feedback energy injected into the galaxies at all times, which increases the star formation rates and the stellar masses over the whole mass range of haloes. The fact that the gas fraction is comparatively higher, particularly at $3\times 10^{12}\, \text{M}_\odot$, is again due to the mass-dependent switch from quasar to the kinetic mode. As the black holes are less massive in the modified run, they mostly remain in the quasar mode, keeping a relatively high gas fraction while the corresponding black holes in the fiducial run have switched to kinetic feedback which lowers the gas fractions by $20-30\%$.


\paragraph*{}The halo mass $M_\text{FOF}$ at which black holes are
seeded has a similar effect: lowering $M_\text{seed}$ and increasing $M_\text{FOF}$ both lead to a delayed black hole growth. In our test
simulation, we place the black holes only in haloes that have grown a
factor of $4$ more in mass compared to our fiducial simulation.  This produces similar
trends, but the effect is much weaker. The delay of the black hole
growth is not as severe as in the previous case, which can be
explained by the accretion rate dependence
$\dot{M}_\text{Bondi}\propto M_\text{BH}^2$, which means that the black holes have
a $16$ times higher accretion rate for the same gas properties but are
seeded in four times more massive haloes. This means that they do not
need a similarly severe change of gas properties as the low mass seeds
to eventually grow into the Eddington limited accretion phase.

All in all, the dependence on the seeding prescription reveals one of
the most important theoretical uncertainties of the black hole
modelling in cosmological simulations. The formation and the early
growth of SMBHs are observationally as well as
theoretically very poorly understood \cite[see][for a review]{2010A&ARv..18..279V}. However, as we just showed, they
have a major impact on the evolution of galaxy properties. One way to
reduce these uncertainties from a simulation point of view is to
constrain the model with observations that also crucially depend on
the seeding and early growth phase, such as the low mass
end of the $M_\text{BH}-\sigma$ and $M_\text{BH}-M_\text{bulge}$
relations, or the abundance of high redshift quasars.

\subsubsection{Other parameter dependences}
\label{subsec:nngb}

\paragraph*{}The reorientation factor $f_{\rm re}$ determines the energy threshold at which a new kinetic feedback event along a new direction is injected. The specific parameterization we adopted sets the magnitude of the velocity kicks relative to the local dark matter velocity dispersion. The remarkable thing about lowering this parameter by a factor of $4$ is that the black hole mass does not change at all, while the star formation rate and correspondingly the stellar masses as well as the gas mass increase significantly. This means that this factor substantially changes the efficiency of the kinetic feedback and therefore the properties of the high-mass haloes. The fact that the burstiness of the feedback has such a dramatic impact is in agreement with other works \citep[e.g.][]{2014MNRAS.441.1270L,
  2015MNRAS.452..575S} and can in our case be explained by the fact that the velocity kicks directly determine the strength of the resulting shocks as well as the post shock temperature. The faster the velocity, the higher the post-shock temperature and the lower the cooling losses during the process. With the adopted parameters, we reach velocity kicks up to several tens of thousand $\text{km\,s}^{-1}$ in the largest haloes, which are realistic speeds for winds from optically thin accretion discs. This means that one could in principle try to constrain this parameter, both theoretically from small-scale GRMHD simulations of hot accretion flows \citep{2015ApJ...804..101Y} as well as from observations \citep{2014MNRAS.443.2154T}.

\paragraph*{}The number of neighbours $n_\text{ngb}$ sets the number of cells used for the density, sound speed and velocity estimates, as well as for the injection region of feedback energy. Increasing this number means that, at a fixed resolution, the radius out to which the gas properties are probed increases. As the gas properties change with radius, the accretion rate estimate tends to change as well. This has important consequences for the black holes in low mass systems, because here the black hole growth is delayed when we average over a four times larger number of cells, which can be seen in Fig.~\ref{fig:BHARD_Z}. This leads to a lower black hole mass for galaxies in $10^{12}\,\text{M}_\odot$ haloes. For more massive haloes, the black hole mass increases by about $25 \%$, and correspondingly lower gas and stellar mass fractions are reached. However, Fig.~\ref{fig:BHARD_Z} also reveals a higher BHARD between redshifts $z=3$ and $0.5$, which is responsible for the more massive black holes. This is because the quasar mode distributes the energy over more mass, leading to lower temperatures in the heated gas and higher radiative cooling losses.  In the kinetic mode, the larger injection volume means that we implicitly increase the burstiness of the model, which increases its efficiency at quenching star formation. This explains the lower star formation rate for the most massive haloes and the lower star formation rate after $z=1$ for haloes in the mass range $10^{12}\,\text{M}_\odot < M_{200,c} < 10^{12.5}\,\text{M}_\odot$.


\section{Conclusions}
\label{sec:Conclusion}

In this study, we introduced a new model for SMBH growth and the associated feedback in cosmological simulations of galaxy formation. We distinguish between a state of high and a state of low accretion, which are associated with pure thermal or pure kinetic feedback, respectively. Unlike in previous work, we omit an artificial boost factor $\alpha$ in the accretion rate estimate to account for unresolved ISM structure, and instead adopt an accretion rate given by the Bondi formula throughout.  The feedback energy in the high accretion rate state is released with a continuous thermal feedback prescription. In the low accretion state, we instead use pulsed kinetic feedback injection in random directions, which is the primary new element adopted in this study. We have shown in idealized simulations that this mode drives shocks in the surrounding gas, thermalizing a significant fraction of the AGN energy within a Myr.

In simulations of cosmological structure formation, our new model is
able to significantly reduce star formation in the most massive haloes,
leading to a stellar mass fraction in excellent agreement with
observations, without overly heating and diluting the central
gas. This resolves one of the central problems in the Illustris
simulation. It also leads to massive galaxies with a
red, old stellar population, living in haloes that have gas fractions in
agreement with observations.

The star formation efficiency peaks in haloes with a few times
$10^{12}\,\text{M}_\odot$, in very good agreement with abundance
matching expectations once we use the halo masses from dark matter
only simulations for the comparison, as also used in the fits to
observations on which the abundance models are based. The key to
sustained quenching of massive haloes in our simulations is to ensure
that the black holes in these systems transition to the low accretion
state and remain in it for most of their subsequent evolution. We encourage
this behaviour by employing a BH mass-dependent Eddington ratio
threshold for determining the accretion state, making it progressively
easier for high-mass black holes to be in the kinetic mode. Once the
black holes reach this mode, the more efficient coupling of the
kinetic feedback and the self-regulated nature of gas accretion will
typically keep the black holes accreting at low Eddington rates. Brief
interruptions of this with episodes of quasar activity, triggered for
example by significant inflows of cold gas during a galaxy merger, may
nevertheless occur.

We analysed the impact of each of our black hole model parameters on
the cosmic star formation rate history and the stellar, gas and black
hole masses. To this end we varied each parameter by a factor of $4$
and carried out otherwise identical simulations to our default model.
We found that most of the parameters do not alter the global
properties severely, but some of them can have a significant impact on
a subset of haloes and galaxies over particular mass ranges. In these
cases, the changes can be readily understood in terms of the tightly
self-regulated nature of black hole growth that occurs in our models. We
would like to emphasize that the assumption of the existence
of a low accretion rate state with efficient kinetic feedback is more important
than the precise value of any of the model parameters.

The new AGN feedback model discussed here significantly improves the galaxy formation model explored previously in the Illustris simulation project, particularly at the high-mass end of the galaxy population. It therefore promises to be an excellent starting point for a new generation of hydrodynamical simulations of galaxy formation that allow much improved predictions for the bright end of the galaxy population, and for groups and clusters of galaxies, as well as their thermodynamic scaling relations.  Future work with this model in high-resolution simulations of galaxy formation could potentially also shed light on the physical origin of observed centrally concentrated radio emission \citep{2015A&A...576A..38B, 2016AN....337..114B}, AGN driven nuclear outflows \citep{2014ApJ...787...38F, 2014MNRAS.443.2154T} and related phenomena.

\section*{Acknowledgements}
The authors thank Peter Behroozi for providing his data and for useful advice, as well as Kevin Schaal for providing his shock finding algorithm.  RW, VS and RP acknowledge support through the European Research Council under ERC-StG grant EXAGAL-308037. RW, VS and RP would like to thank the Klaus Tschira Foundation. RW acknowledges support by the IMPRS for Astronomy and Cosmic Physics at the University of Heidelberg. SG and PT acknowledge support provided by NASA through Hubble Fellowship grant HST-HF2-51341.001-A and HST-HF2-51384.001-A, respectively, awarded by the STScI, which is operated by the Association of Universities for Research in Astronomy, Inc., for NASA, under contract NAS5-26555. LH acknowledges support from NASA grant NNX12AC67G and NSF grant AST-1312095.  Simulations were run on the HazelHen supercomputer at the High-Performance Computing Center Stuttgart (HLRS) as part of project GCS-ILLU of the Gauss Centre for Supercomputing (GCS).




\bibliographystyle{mnras}




\appendix

\section{Changes in the wind model}
\label{app:wind}

\begin{figure}
\includegraphics{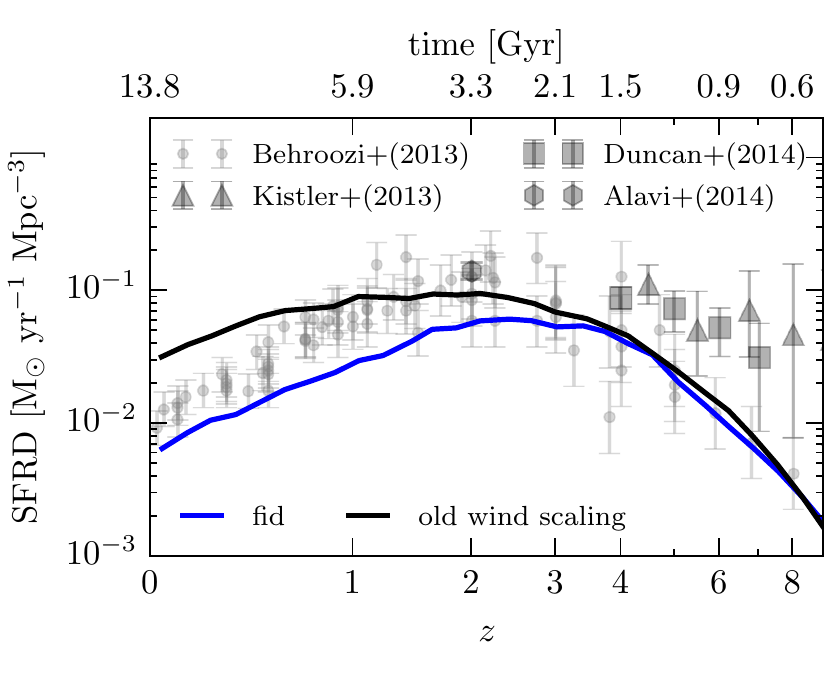}
\caption{SFRD versus redshift for the fiducial run
  and a run with the old wind scaling.}
\label{fig:SFRD_Z_wind}
\end{figure}

In addition to the changes in the AGN model, we have implemented some alterations to the stellar wind feedback compared to the model described in \citet{2013MNRAS.436.3031V}. They are introduced to address some of the shortcomings of the Illustris simulation in low-mass systems, such an excessive number of galaxies with blue star forming rings, a too high stellar mass fraction in systems with halo masses $M_{200,c} < 10^{11.5}\, \text{M}_\odot$ and a too mild decline in SFRD at low redshift \citep{2014MNRAS.444.1518V}. We summarize these changes here for completeness and refer to Pillepich et al.~(in preparation) for a detailed discussion. 

First, we now use an isotropic wind injection with $10\%$ of the energy injected thermally and not the bipolar, purely kinetic approach employed in the Illustris project \citep{2014Natur.509..177V}. 
Furthermore, we slightly changed the wind velocity. We still use the scaling with local dark matter velocity dispersion as in equation 14 of \citet{2013MNRAS.436.3031V}, but introduce an additional redshift dependent factor $[H_0 / H(z)]^{1/3}$, which effectively yields a scaling of the wind velocity that depends purely on halo mass. Additionally, we set a minimum wind velocity of $v_\text{min} = 350\,\text{km}\,\text{s}^{-1}$ to prevent unrealistically high mass-loading factors in low-mass haloes. Taken together, the equation for the wind velocity is hence 
\begin{align} 
v_w = \max\left[ \kappa \,\sigma_\text{DM}^\text{1D} \left(H_0/H(z)\right)^{1/3}, v_\text{min} \right]. 
\end{align}

We choose the parameters such that the wind velocity of a given halo equals that of the previous Illustris model at a redshift $z\simeq 5$, implying that it then tends to increase slightly towards lower redshifts compared to \citet{2014MNRAS.444.1518V}.  This is the main reason for the different scaling of the SFRD with redshift (Fig.~\ref{fig:SFRD_Z_wind}). The minimum wind velocity is partially responsible for the sharp decline in star formation efficiency towards lower masses in Fig.~\ref{fig:fStarICL_Mhalo}.  A summary of the adopted wind parameters and a comparison to those used in Illustris is given in Table~\ref{tabwinds}.

Moreover, we use a higher baseline wind energy for gas of primordial abundance but now reduce the available energy with metallicity $Z$ on the grounds that higher metallicity galaxies plausibly have larger radiative cooling losses of the supernova energy. A similar factor has also been used in the Eagle project \citep{2015MNRAS.446..521S}.  The energy of the winds is reduced by a factor \begin{align}
 f + (1 - f) / \left[ 1 + \left({Z}/{Z_\text{red}}\right)^{\gamma} \right]
\end{align}
where $f = 0.25$, $\gamma = 2$ are free parameters and
$Z_\text{red} = 0.002$.  This effectively lowers the efficiency of the
supernova feedback in metal-enriched galaxies, and since most of the stars form there, the total injected wind energy is comparable to Illustris. However, the metal dependence leads to
a higher relative efficiency of the wind feedback in low-mass systems,
suppressing them more in comparison to Milky Way-sized galaxies, which
is an effect that seems required by the observational data and the low
abundance of luminous dwarf galaxies.
\begin{table}
\begin{tabular}{c c c r}
  \hline
 Parameter & New & V13 & Note\\
\hline
 $\kappa$ & 7.4 & 3.7 & Same velocity at $z\approx 5$\\
 egy$_w$ /egy$_{0w}$ & 3.6 & 1.09 & Reduced through metallicity\\
          &     &     & Dependence\\
 
\hline
\end{tabular}
\caption{Comparison of wind parameters with \citet{2013MNRAS.436.3031V,2014MNRAS.438.3607V} (V13). \label{tabwinds}}
\end{table}

All in all, our stellar feedback is somewhat stronger than in the
Illustris simulation \citep{2014MNRAS.444.1518V}, particularly at late
times and in low-mass haloes. At high redshift, the strength of the
stellar feedback is comparable, which also means that the $z=0$ black
hole masses are not significantly affected by the changes.

\section{Numerical convergence}
\label{app:convergence}

\begin{figure}
\includegraphics{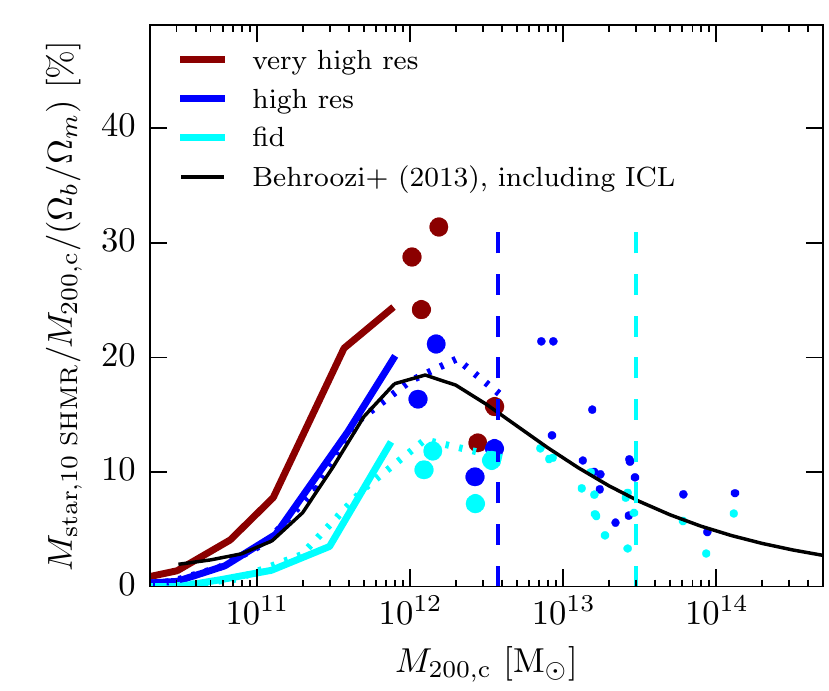}
\caption{Stellar mass fraction as a function of halo mass for
  simulations of different resolution. The dotted lines and small dots
  indicate the simulations with $30 \, h^{-1} \text{Mpc}$ side length,
  the solid lines and large dots test simulations with
  $7.5\,h^{-1} \text{Mpc}$ side length. Note that the colour coding
  is different compared with results presented in the main text. The dashed vertical
  lines are at $5\times10^5\,(m_\text{gas} + m_\text{dm})$, i.e. these
  haloes would have $5\times10^5$ simulation particles within
  $R_\text{200,c}$ if their baryon content was equal to the cosmic baryon
  fraction.}
\label{fig:fStar_convergence}
\end{figure}

Achieving numerical convergence is a major challenge for full physics
cosmological volume simulations due to the multi-scale, multi-physics
nature of the problem. Normally, convergence cannot be fully
established, and hence the lack thereof represents an additional
source of systematic uncertainty in the predictions.  We attempt to
quantify the magnitude of resolution effects here, using simulations with a
box side length of $7.5 \, h^{-1} \text{Mpc}$ with
$2\times384^3, 2\times192^3$ and $2\times96^3$ simulation particles
and cells (dark matter + gas). In this small simulation
volume, only low-mass haloes form, which makes an analysis of the
global SFRD and BHARD meaningless, but it still allows us to estimate the
uncertainties due to numerical convergence for the galaxies that
happen to be present, especially because we can simulate them at
resolutions higher than our standard high-resolution simulation with
box side length of $30\,h^{-1}\text{Mpc}$. 

{We focus on the bias due to resolution effects at the low-mass end, as this regime has been the most severely affected when increasing the resolution in the Illustris simulation \citep[][fig. A1, upper panel]{2014MNRAS.444..237P}. This is not surprising as the {\small AREPO} code ensures that the individual gas cells have approximately equal mass and consequently the number of gas cells within a halo decreases rapidly with decreasing halo mass. Fig.~\ref{fig:fStar_convergence} shows the star formation efficiency as a function of halo mass for different resolutions. The vertical dashed lines correspond to $5\times 10^5 (m_\text{gas}+m_\text{dm})$, which, depending on the gas fraction in the halo, translates to a few times $10^5$ gas cells within $R_{200,c}$. Decreasing the number of gas cells in a halo, individual cells} become so large that they average over significant regions of the ISM, producing lower average densities and hence longer gas consumption time-scales. This results in a numerically suppressed star formation rate and, over time, in a lower stellar mass fraction. As this convergence issue is present for the haloes at the peak of the star formation efficiency, it also manifests itself in the global SFRD.

A second resolution problem, related to black holes, is the radius $h$ over which the gas properties are averaged to derive an accretion rate estimate and to inject the feedback energy. We adjust this radius such that it contains approximately a constant number of cells $n_\text{ngb}$. In Section~\ref{sec:parameters}, we presented the effect of increasing $n_\text{ngb}$ by a factor of $4$. Changing the particle number per dimension by a factor of $2$, $h$ also decreases, assuming constant $n_\text{ngb}$. This means that the volume over which the gas properties are averaged is a smaller volume at the centre of the galaxy and therefore usually denser, which leads to higher accretion rate densities, as seen in Fig.~\ref{fig:BHARD_Z}. If we increased $n_\text{ngb}$ to keep approximately the same volume to average over, this effect would be smaller, but we would not benefit from the increased spatial resolution in the centre. For our simulation sequence, we aimed for a compromise by increasing $n_\text{ngb}$ by a factor of $2$ whenever the particle number per dimension is increased by a factor of $2$.


\bsp	
\label{lastpage}
\end{document}